\documentclass[11pt, a4paper]{article}

\usepackage{authblk}
\usepackage[round]{natbib}
\usepackage{graphicx}        % For eps figures, newer & more powerfull
\usepackage{color}           % For color text: \color command
\usepackage{breakurl}        % For breaking URLs easily trough lines
\usepackage[outdir=./]{epstopdf}
\usepackage{lineno}
\usepackage{soul}

\usepackage[citecolor=blue,
            anchorcolor = blue,
            colorlinks=true,
            linkcolor = blue,
            urlcolor=blue,
            breaklinks=true]{hyperref} 
\chardef\us=`\_

\title{\textbf{Simulating White-Light Images of Coronal Structures for Parker Solar Probe/WISPR: Study of the Total Brightness Profiles}}

\author[1,2]{Giuseppe~Nistic\`o}%sep
\author[1]{Volker~Bothmer}%\sep
\author[3]{Angelos~Vourlidas}%\sep
\author[4]{Paulett~Liewer}%\sep
\author[5]{Arnaud~Thernisien}%\sep
\author[5]{Guillermo~Stenborg}%\sep
\author[5]{Russell~Howard}
\affil[1]{\small{\href{mailto:giuseppe.nistico.fis@unical.it}{giuseppe.nistico.fis@unical.it}, Institut f\"ur Astrophysik, Georg-August Universit\"at, G\"ottingen, 37077, Germany}}
\affil[2]{\small{Department of Physics, University of Calabria, Ponte P. Bucci, Cubo 31 C, 87036 Rende, Italy}}
\affil[3]{\small{Applied Physics Laboratory, Johns Hopkins University, Laurel, MD 20723, USA}}
\affil[4]{\small{Jet Propulsion Laboratory, California Institute of Technology, Pasadena, Ca 91109, USA}}
\affil[5]{\small{Naval Research Laboratory, Washington, DC 20375, USA}}

\date{}

\begin{document}

\maketitle
%\runningauthor{Nistic\`o et al.}
%\runningtitle{Simulating images for PSP/WISPR: study of total brightness profiles}

\begin{abstract}
The Wide-field Imager for Parker Solar Probe (WISPR) captures unprecedented white-light images of the solar corona and inner heliosphere. Thanks to the uniqueness of Parker Solar Probe's (PSP) orbit, WISPR is able to image ``locally'' coronal structures at high spatial and time resolutions. The observed plane of sky, however, rapidly changes because of the PSP's high orbital speed. Therefore, the interpretation of the dynamics of the coronal structures recorded by WISPR is not straightforward. 
A first study, undertaken by \citet{Liewer2019}, shows how different coronal features (e.g., streamers, flux ropes) 
appear in the field of view of WISPR by means of raytracing simulations. In particular, they analyze the effects of the spatial resolution changes on both the images and the associated height-time maps, and introduce the fundamentals for geometric triangulation. In this follow-up paper, we focus on the study of the total brightness of a simple, spherical, plasma density structure, to understand how the analysis of Thomson-scattered emission by the electrons in a coronal feature
can shed light into the determination of its kinematic properties. We investigate two cases: {\it (a)} a density sphere at a constant distance from the Sun for different heliographic longitudes; {\it (b)} a density sphere moving outwardly with constant speed. 
The study allows us to characterize the effects of the varying heliocentric distance of the observer and scattering angle on the total brightness observed, which we exploit to contribute to a better determination of the position and speed of the coronal features observed by WISPR.

\end{abstract}
%-------------------------------------------------

\section{Introduction}
     \label{S-Introduction} 
     
The launch of Parker Solar Probe \citep[PSP,][]{Fox2016} has raised a lot of excitement and expectations in the science community. In addition to the in-situ instruments \citep{Bale2016,Kasper2016,McComas2016}, which make direct measurements of the local environment, PSP is equipped with the Wide-field Imager for Parker Solar Probe \cite[WISPR;][]{Vourlidas2016}. This unique, white-light imaging telescope suite images the solar corona and inner heliosphere starting at $13.5 \deg$ elongation, from an heliocentric distance of a few tens of solar radii from the Sun's centre.
%the photosphere.

Similarly to the heliospheric imagers \citep{Howard2008} aboard the Solar Terrestrial Relations Observatory \citep[STEREO,][]{Kaiser2008}, WISPR consists of two cameras: the inner telescope (hereafter WISPR-I) with a field-of-view (FoV) of $40 \deg$ and optical axis offset from the Sun centre by $33.5 \deg$; and the outer telescope (hereafter WISPR-O) with a FoV of $58 \deg$ and optical axis offset by $79 \deg$. WISPR is mounted on the spacecraft (S/C) ram direction, therefore plasma structures are observed in advance and eventually crossed by PSP and measured in-situ by the other instruments. 

Both imagers record the total brightness emission due to 1) Thomson scattered light by free electrons in coronal features and solar wind (K-corona), and 2) scattered light by dust particles in orbit around the Sun (F-corona). The latter component is the dominant signal and must be removed to reveal the fainter K-corona signal \citep[][]{Howard2019}.
The fast transit of PSP at the perihelia (PSP will move faster than 200 km s$^{-1}$ in 2024) poses novel challenges on the interpretation and analysis of WISPR images. Contrarily to usual white-light imagers (e.g. those on STEREO where observations are performed from near 1 AU), WISPR data are affected by a rapid change of the plane-of-sky (PoS) and spatial resolution, resulting in significant varying conditions of the background emission. These effects inevitably influence the physical characterisation of the observed coronal structures, as well as the image calibration and processing \citep{Vourlidas2016}.

\begin{figure}[htpb]
	\centering
	\includegraphics[width = .6\textwidth]{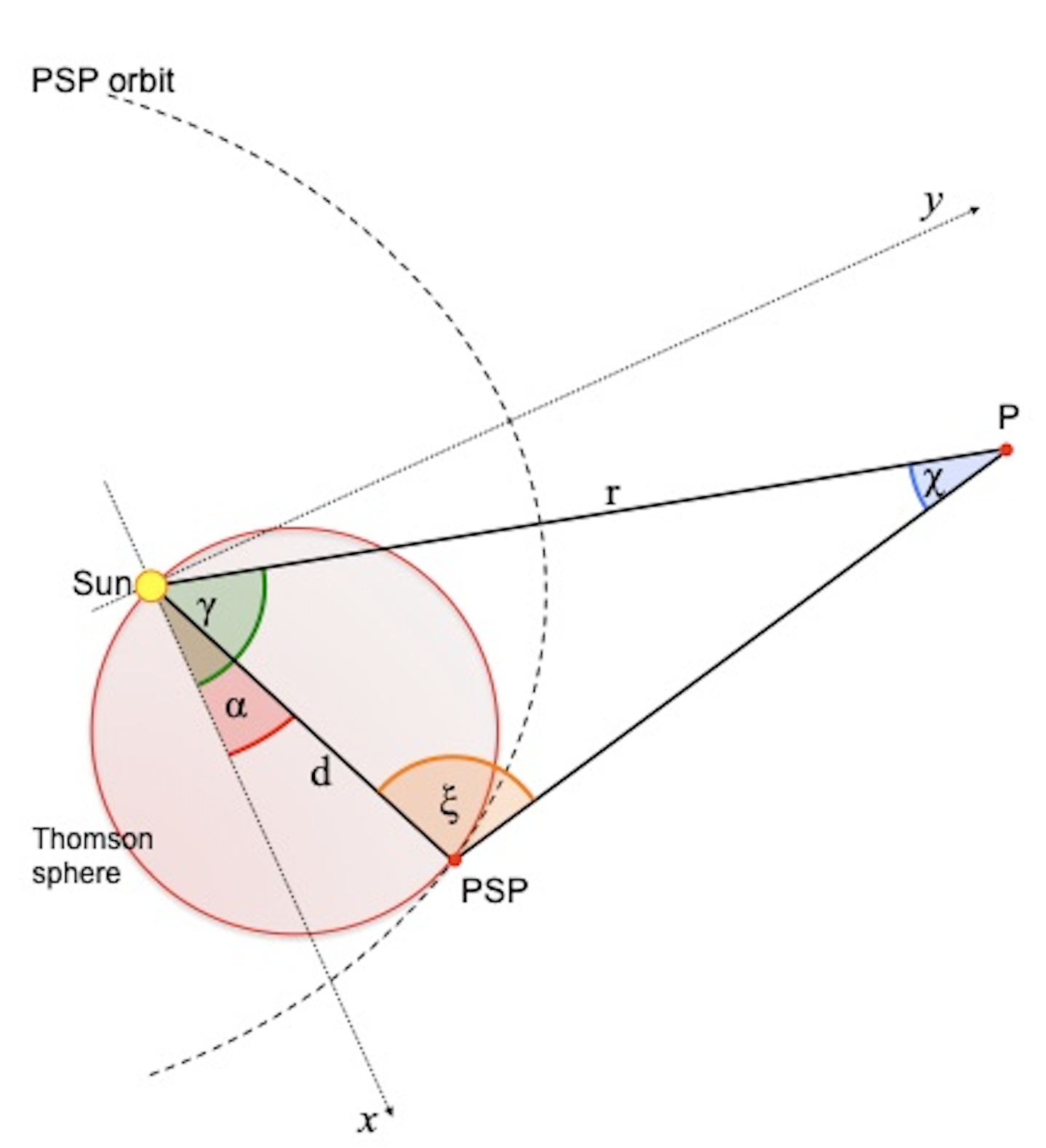}
    	\caption{Two-dimensional representation of the Thomson scattering geometry in a reference frame defined by the plane $xy$, coincident with the orbital plane of PSP, and centred on the Sun: the scattering feature P has polar coordinates $r$ and $\gamma$, PSP has coordinates $d$ and $\alpha$ and its orbit is shown as a dashed line. PSP observes P at an elongation angle $\xi$. The scattering angle in P is defined as $\chi$. The red circle represents the Thomson sphere.}
	\label{fig_sketch}
\end{figure}

\citet{Vourlidas2006} show that the Thomson scattered emission from a feature, like a coronal mass ejection (CME), depends on the geometry between the observer and the scattering structure. Indeed, they demonstrate that, along a given line of sight (LOS), the maximum of scattering emission is on a sphere, defined as the Thomson sphere (TS), with a diameter equal to the Sun-observer distance \citep[see Fig. 1 of][]{Vourlidas2006, Vourlidas2016}. The TS is widely accepted to represent a reference surface for understanding the variations of the total brightness. Certainly, in the case of PSP, the continuous change in distance from the Sun modifies the extension of the TS, as well as the location of a coronal structure with respect to it (Fig. \ref{fig_sketch}). 

The mathematical expression for the total brightness requires the evaluation of an integral along the LOS. In the approximation of large distances from the Sun, the mathematical formula is given by Eq. 4 in Sect. 2 of \citet[][]{Howard2012}. For simplicity, we can consider that the average total brightness per pixel, measured in units of $\mathrm{W m^{-2} sr^{-1}}$ or mean solar brightness (MSB), of an extended feature moving radially along the orbital plane of the S/C with polar coordinates $(r(t),\gamma)$ can be expressed as
%The total brightness $B$ of an observed feature depends on its electron number density content $n$, the column depth along the line-of-sight $h$, and its distance from the Sun $r$ (Fig. \ref{fig_sketch}) as
	\begin{equation}
		B(t) = \frac{C}{r(t)^2} (1+\cos^2\chi(t)),
		\label{eq_tb}
	\end{equation}
with $r(t)$ the radial distance from the Sun, $\chi(t)$ the scattering angle determined by the direction of the incident radiation to the structure and the LOS. The factor $C$ is equal to $C=\pi B_\odot R^2_\odot \sigma_t  n h$ \citep{Howard2012} and includes: the Sun's radiance ($B_\odot\approx2.3\times10^7~\mathrm{W m^{-2} sr^{-1}}=1~\mathrm{MSB}$), the Sun's radius ($R_\odot$), the differential Thomson cross-section \citep[$\sigma_t$, defined for perpendicular scattering according to][]{Howard2009}, the plasma density number ($n$) and the average column depth of the feature ($h$).
%
%	\begin{equation}
%	k = \pi \sigma_e B_\odot  R_\odot^2/2 \approx 2.8 \times 10^{-4}~\mathrm{W m^2 sr^{-1}} = 1.2 \times 10^{-11}~\mathrm{MSB~m^4}. 
%	\label{eq_k}
%	\end{equation}	
The structure is located on the TS when $\chi= \pi/2$. Given the 2D geometric configuration in Fig. \ref{fig_sketch} (as a first approximation we assume that the orbital plane of the spacecraft coincides with the solar equatorial plane), it is trivial to show that $\chi(t)=\pi - \left[ \gamma - \alpha(t) + \xi(t) \right]$, hence the total brightness can be rewritten as a function of the elongation angle $\xi(t)$ as

	\begin{equation}
		B(t) =  \frac{C}{d(t)^2 \sin^2 \xi(t)} \left[1 - \cos^4(\gamma - \alpha(t) - \xi(t))\right],
		\label{eq_tb_xi}
	\end{equation}
with $\gamma$ the only feature's position coordinate, $d(t)$ and $\alpha(t)$ functions of time for the distance and the azimuthal angle (measured in the direction of the orbital motion from any reference axis) of PSP. Indeed, Eq. \ref{eq_tb_xi} could be ideally used for the determination of $\gamma$ from a set of measurements of elongation and total brightness $\{\xi_i, B_i\}$. However, the rapid variation of PSP's position implicates some difficulty in using Eq. \ref{eq_tb_xi} as a fitting function (e.g., for observers like SoHO or STEREO, $d$ and $\alpha$ would be almost constant).

On the other hand, the analysis of the total brightness profiles on WISPR images provides useful information on the observed structure. For example, if a feature moves with constant speed $v$, keeping its size and density constant, the total brightness would decrease according to Eq. \ref{eq_tb} via the variation of the radial distance $r$ and the scattering angle $\chi$, and to Eq. \ref{eq_tb_xi} via the variation of $\xi$, which is directly measured from observations. Therefore, one of these equations could be used to infer the propagation speed of the structure. Indeed, the quantification of the speed via photometric analysis could additionally constrain the determination of the 3D location of the structure. The basics for triangulation and 3D reconstruction with WISPR are already described in \citet{Liewer2019}. 

In this paper, we investigate the total brightness profiles as a function of time for a coronal structure observed with WISPR, by creating synthetic images using the raytracing code distributed within the IDL/Solarsoftware (SSW) library \citep[][]{Thernisien2006}. We consider a simple geometric structure, that is a sphere of plasma, which mimics the appearance and propagation of CMEs, 
%\st{\citep{Viall2015}}
and analyse two different cases for PSP/WISPR observations: (a) a density sphere of constant size at a constant heliocentric distance $r$ and different heliocentric longitudes; and (b) a density sphere of constant size moving away from the Sun. The paper is structured as follows: Sect. \ref{sect2} describes the raytracing software and the developments made to adapt the code to the PSP orbit and the WISPR FoV; Sect. \ref{sect3} and \ref{sect4} present the above mentioned cases for an ideal stationary observer and PSP, respectively; Sect. \ref{sect5} discusses the triangulation of the features with WISPR. Remarks and conclusions are finally presented in Sect. \ref{sect6}. 

\section{Raytracing software: adaptation to the PSP orbit}
\label{sect2}

The raytracing software developed by  \citet{Thernisien2006} simulates the Thomson scattering emission of a coronal structure as viewed from an observer. The full code is available in SSW \footnote{\url{https://sohowww.nascom.nasa.gov/solarsoft/stereo/secchi/cpp/scraytrace/}}.
%A full description of the code and examples are available online {\footnote{\url{https://secchi.nrl.navy.mil/synomaps/scraytrace/doxy/index.html}}\footnote{\url{https://secchi.nrl.navy.mil/synomaps/scraytrace/dobo/index.html}}}.
The software is particularly relevant in the context of CMEs and the graduated cylindrical shell (GCS) forward modelling \citep{Thernisien2006,Bosman2012}. Moreover, it implements raytracing for various analytically-defined coronal structures.
%For simplicity, as mentioned in the previous section, in our tests we considered a simple plasma sphere, that is model 58 of the raytracing code.

\begin{figure}[htpb]
	\centering
		\includegraphics[width=1.0\textwidth]{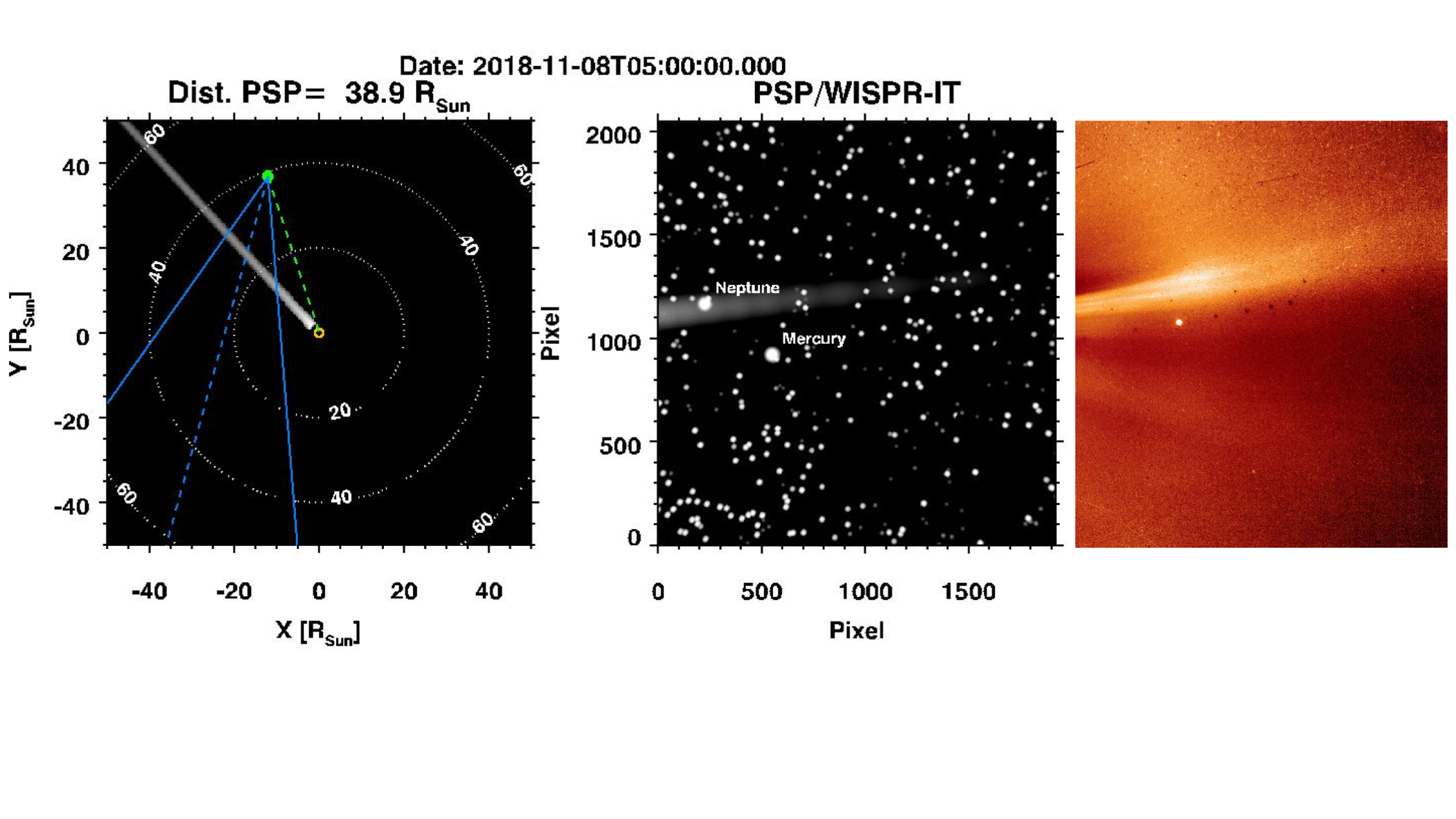}
		\caption{Example of output of the widget showing the PSP trajectory and a coronal streamer (left), and the synthetic view of WISPR-I (middle). The FoV of WISPR-I is represented by the blue cone. Associated animation online in \href{run:movie/movie1.mp4}{movie1.mp4}. A background-corrected WISPR-I image obtained during the first perihelion of PSP is shown for comparison (right). Courtesy of NASA/NRL/Parker Solar Probe.}
		\label{fig_widget}
\end{figure}

To simulate the WISPR perspective, we had to take into account the PSP trajectory and the pointing of the WISPR instruments. Such information is typically available in the header of a FITS file of any solar image data. Therefore, our strategy was to simulate a WISPR FITS header and passed it as an input parameter to the raytracing routine \texttt{rtraytracewcs.pro}.
To define a WISPR header, first we calculated the PSP position at a desired time from the latest SPICE kernel (available in SSW in the \texttt{psp} directory) by using the routines of the SUNSPICE package (documentation available in SSW\footnote{\texttt{ssw/packages/sunspice/doc/sunspice.pdf}}), specifically with the function \texttt{get\_sunspice\_lonlat.pro}. Then, given the position of PSP, we built a World Coordinate System (WCS) structure via \texttt{wcs\_2d\_simulate.pro}, where we defined all the basic and necessary information for a WISPR image: i.e. image size, radial distance and heliographic coordinates of the observer, pixel units, and image centre. Finally, the FITS header was built from the WCS structure, via the routines \texttt{wcs2fitshead.pro}, \texttt{fxaddpar.pro} (to add further tag names), and \texttt{fitshead2struct.pro}.
The overall steps just described are organised in a widget program called \texttt{wisprraytrace}, which allows the user to select a given time interval, time cadence for a sequence of images, and to plot a view of PSP's trajectory and the location of the desired coronal structure, as well as the synthetic WISPR image. The raytracing software allows for several analytical structures to be modelled and simulated, like a sphere for density blobs, cylinders or funnels for streamers, and the graduated cylindrical shell (GCS) model for flux ropes. An example of its output is given in the left and middle panels of Fig. \ref{fig_widget} for the first perihelion in Nov 2018 and compared with an image of the WISPR-I\footnote{\url{https://www.nasa.gov/feature/goddard/2018/preparing-for-discovery-with-nasas-parker-solar-probe} telescope} (right panel).  The left panel shows a view above the solar equatorial plane with a streamer extending over 30 solar radii (model number 17 of the raytracing software), the position of PSP (green) and the the angular FoV (blue) of WISPR-I. The middle panel presents the synthetic image with the appearance of a streamer over the background with stars and the location of the planets Neptune and Mercury. The simulated scenario is qualitatively similar to that presented in the actual WISPR image, which shows an extended streamer and the planet Mercury as a bright spot in the middle of the image. The absence of stars and the presence of dark spots are due to background corrections.

The synthetic images presented hereinafter are not created with the nominal size of WISPR data, i.e. 1920$\times2048$ pixels, but ten times smaller (192$\times$205 pixels) since raytracing calculations are time-expensive. The raytracing code is further adapted to the case of a stationary observer, which is useful for a better understanding and comparison with PSP. The full code is available online on the Zenodo platform\footnote{\url{https://doi.org/10.5281/zenodo.3478623}}. 

 \section{Simulations for a stationary observer}
\label{sect3}

\begin{figure}[htpb]
		\centering
			\includegraphics[width=1.\textwidth]{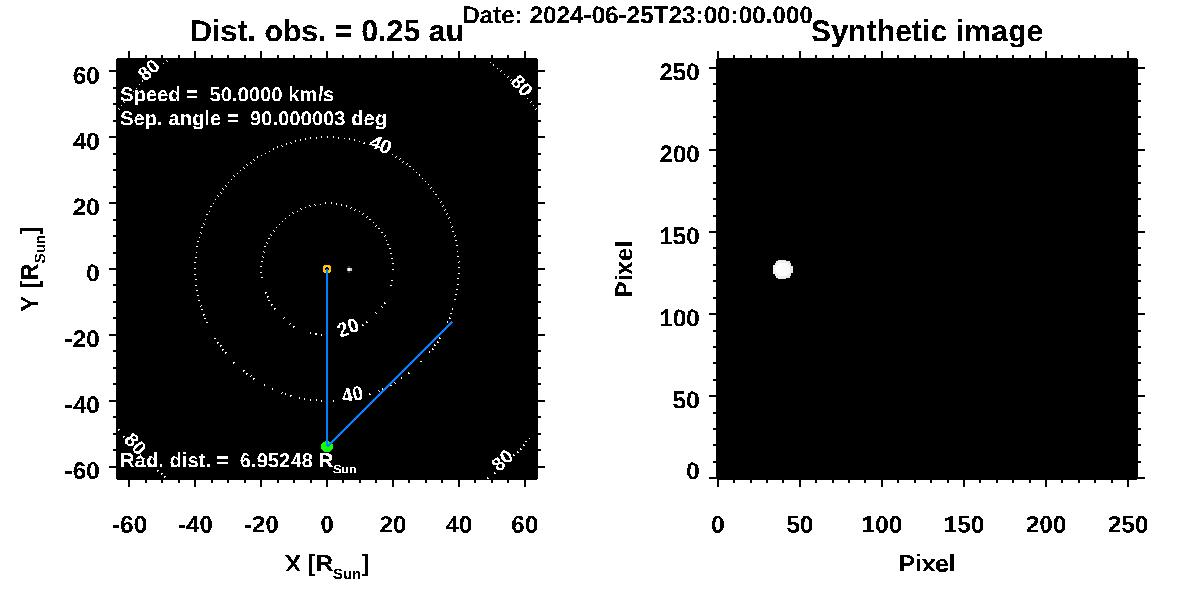} 
			
			\caption{Example of output of the widget showing the observer as a green dot, distant 0.25 au from the Sun, and a density sphere travelling with a speed of 100 km s$^{-1}$ perpendicularly to the Sun-observer line and imaged when it is about 13 solar radii from the Sun (left). The FoV is represented by the blue cone. The synthetic view is shown in the right panel. The simulation date is arbitrary. Associated animation online in \href{run:movie/movie2.mp4}{movie2.mp4}.}
	\label{fig_widget_fo}
\end{figure}

We first simulated the Thomson scattering emission by a density sphere measured from an observer at rest in the heliocentric inertial (HCI) coordinate system. An example is given in Fig. \ref{fig_widget_fo}.
The observer is located along the negative direction of the $y$-axis and observing with a FoV of $45 \deg$, with no offset from the Sun. We analysed the two following cases.

\subsection{Density sphere of fixed size moving with constant speed}

%The blob moves 
Here we analyse the blob moving along the positive direction of the $x$-axis. It departs from 1 solar radius, i.e. from the photosphere, with a size $R=1~\mathrm{R}_\odot$ and an electron number density $n_0 = 4.23 \times10^6 \mathrm{cm}^{-3}$. Given this size and density, the total mass would be equal to $\approx 10^{13} \mathrm{kg}$, which can be regarded as an upper limit value for CME mass \citep{Colaninno2009,Bein2013}. We performed some tests for different distances of the observer from the Sun (e.g., Fig. \ref{fig_widget_fo} shows the observer distant 0.25 au from the Sun), but keeping the same FoV of $45 \deg$. The simulated images have a size of 256$\times$256 pixels in order to have the structure sufficiently resolved in the synthetic images even for large distances of the observer ($\geq 0.75$ au). 

	\begin{figure}[htpb]
%		\centering
		\begin{tabular}{c c}
			\includegraphics[trim=2 2 1 2,clip,width=0.466\textwidth]{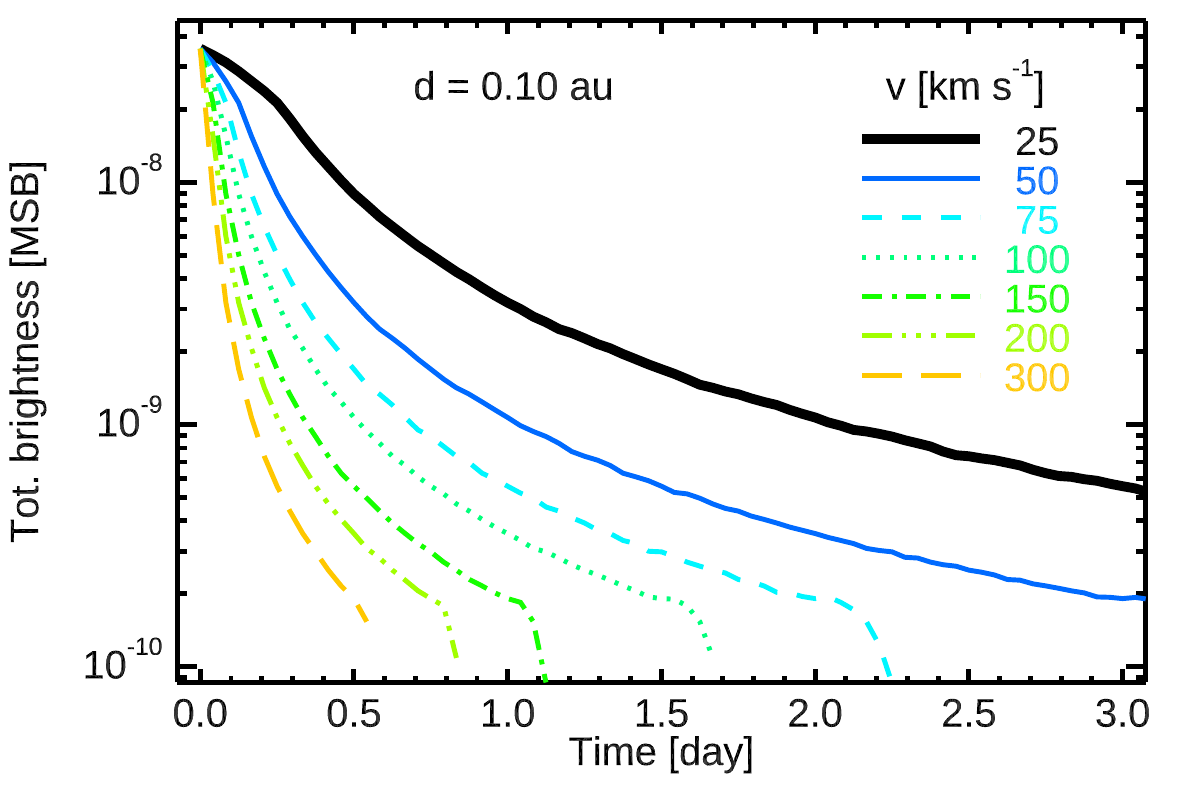} &
			\includegraphics[trim=2 2 1 2,clip, width=0.466\textwidth]{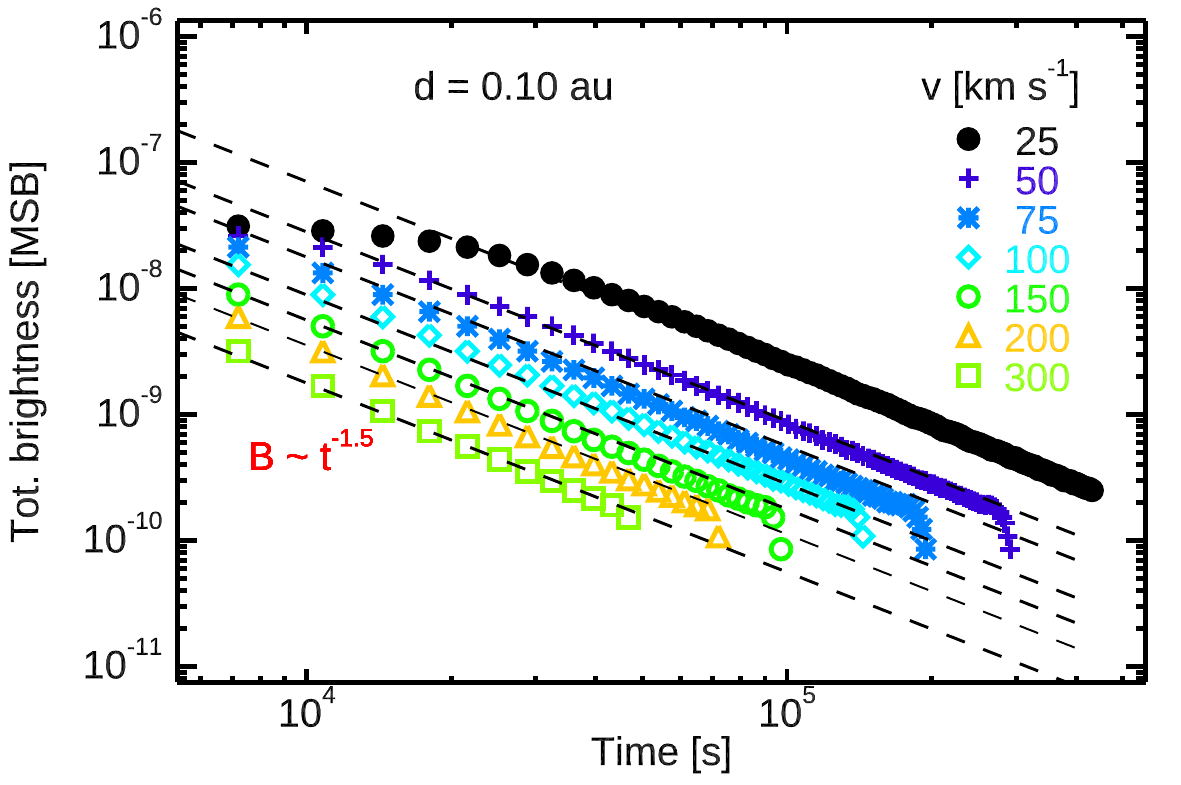} \\
			\includegraphics[trim=2 2 1 2,clip,width=0.466\textwidth]{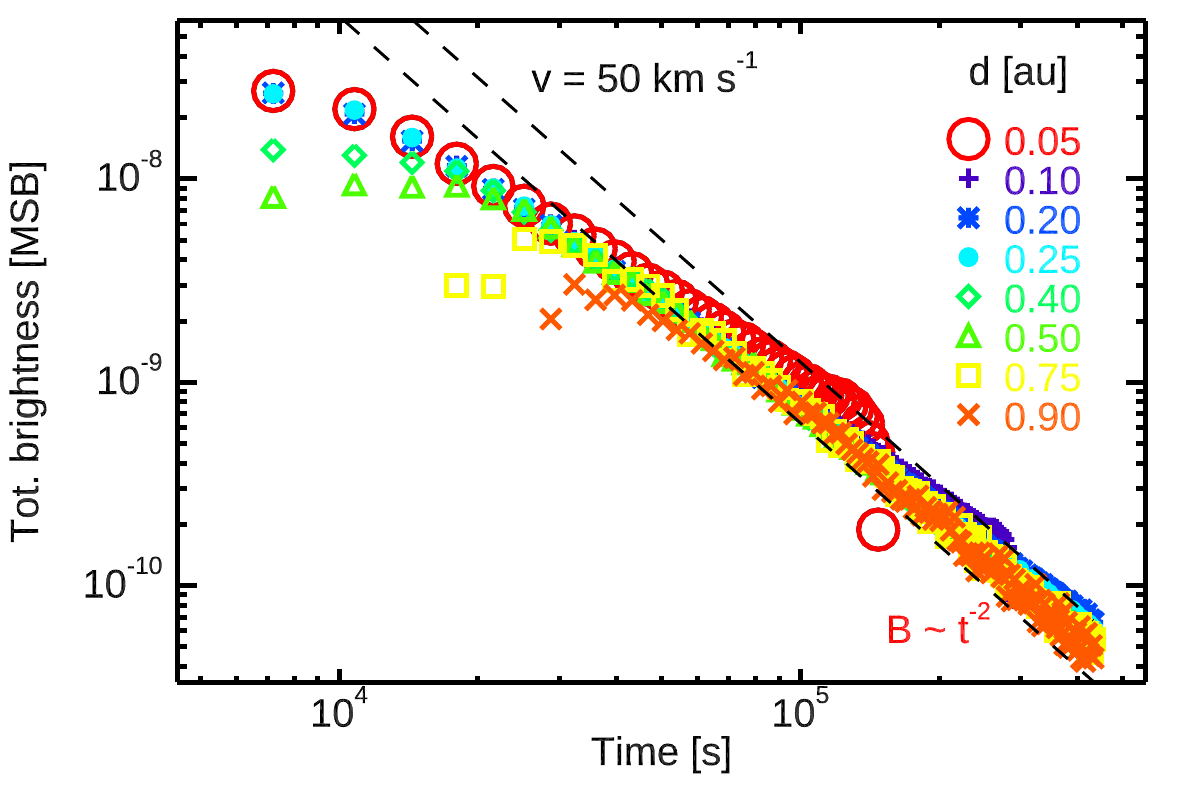} &
			\includegraphics[trim=2 2 1 2,clip,width=0.466\textwidth]{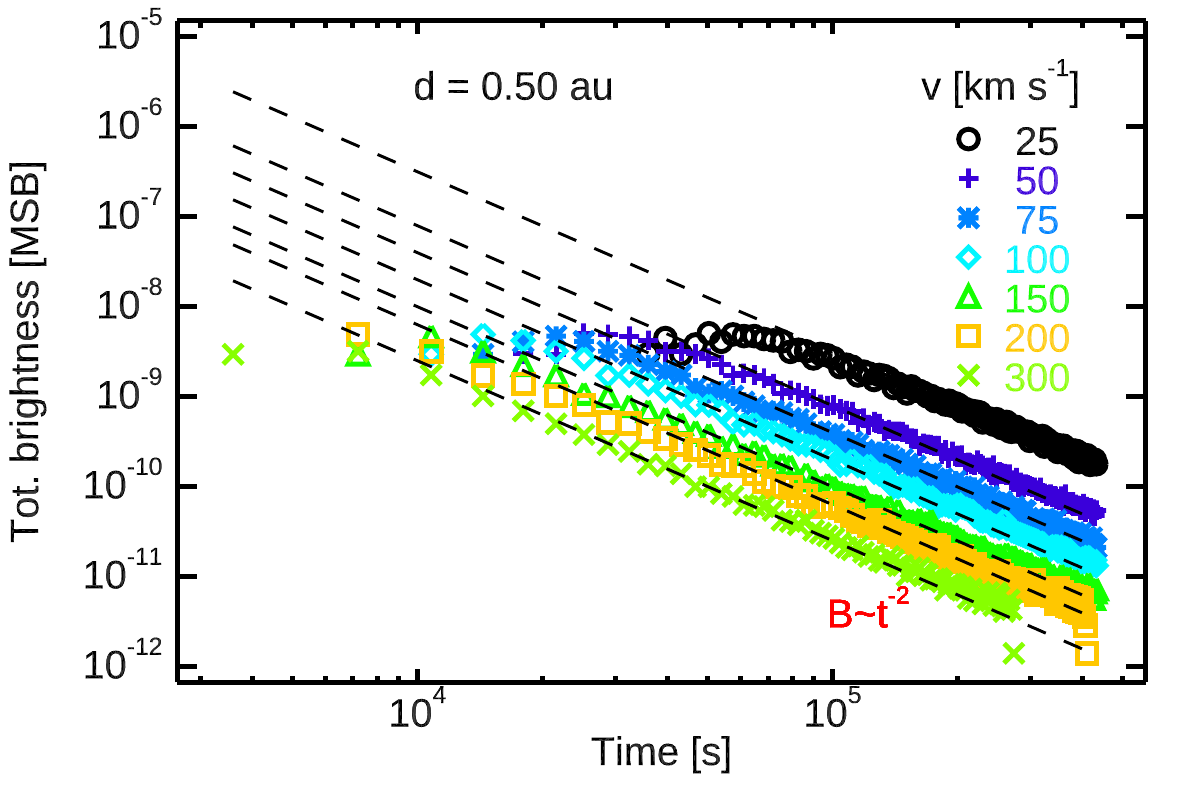} \\
		\end{tabular}
		\caption{Top: average total brightness profiles for a plasma sphere moving with different speeds and observed from 0.10 au in semi-log (left) and log-log (right) scales. Bottom: average total brightness profiles of the plasma sphere moving with a speed of 50 km s$^{-1}$, but observed from different distances (left), and observed from 0.50 au but with different propagation speeds (right).}
		\label{fig_fo_speed}
	\end{figure}
	
To study the evolution of the Thomson scattered emission in the synthetic images, we considered the average total brightness of the structure. In practice, for any synthetic image we selected the pixels with intensity values greater than 0, which belong to the structure (the background is null), and calculated the arithmetic mean of the emission. In the top-left panel of Fig. \ref{fig_fo_speed} we show the average total brightness per pixel as a function of time for the structure moving uniformly with different speeds. The profiles are measured for an observer located at 0.1 AU. The vertical axis is in logarithmic scale. The brightness profiles are trivially steeper for higher speed values, since the sphere gets more distant from the Sun.
	 
If we neglect the contribution of the scattering angle $\chi$, the heliocentric distance $r$, which changes uniformly ($r = r_0 + v t$), affects the total brightness (Eq. \ref{eq_tb}) scaling as  $B \sim r^{-2} \sim t^{-2}$. 
The power law index can be inferred by log-log scale plots of the brightness as a function of time (right panel of Fig. \ref{fig_fo_speed}). In the case of observations taken from 0.1 au, the total brightness scales as $B \sim t^{-1.5}$ (because of the non-negligible contribution of the scattering angle $\chi$). We retrieve the scaling law $B \sim t^{-2}$ at longer times from the instant of ejection of the structure, i.e., when the structure is sufficiently far from the Sun. For example, the bottom-left panel of Fig. \ref{fig_fo_speed} shows the total brightness profiles measured from different distances (0.05, 0.10, 0.25,... au) with the object moving with the same speed of 50 km s$^{-1}$. The initial part of the profiles is flatter because the structure is observed when it is still close to the Sun. This is even more accentuated when the observations are taken from relatively closer distances (0.05-0.50 au). The bottom-right panel presents a similar scenario. However, the comparison is made for brightness profiles of a density sphere moving with different speeds but observed from the same distance (0.5 au). These considerations, despite being trivial, emphasise how the brightness evolution may be related to the kinematics of the observed feature. This scenario is quite simple (absence of background in the images, no expansion and/or density variations in the structure), but it is nevertheless useful for the analysis of WISPR data.

\subsection{Density sphere launched at different heliocentric longitudes}
	
\begin{figure*}[htpb]
		\centering
		\begin{tabular}{c c}
			\includegraphics[trim=3 2 1 2,clip,width=0.465\textwidth]{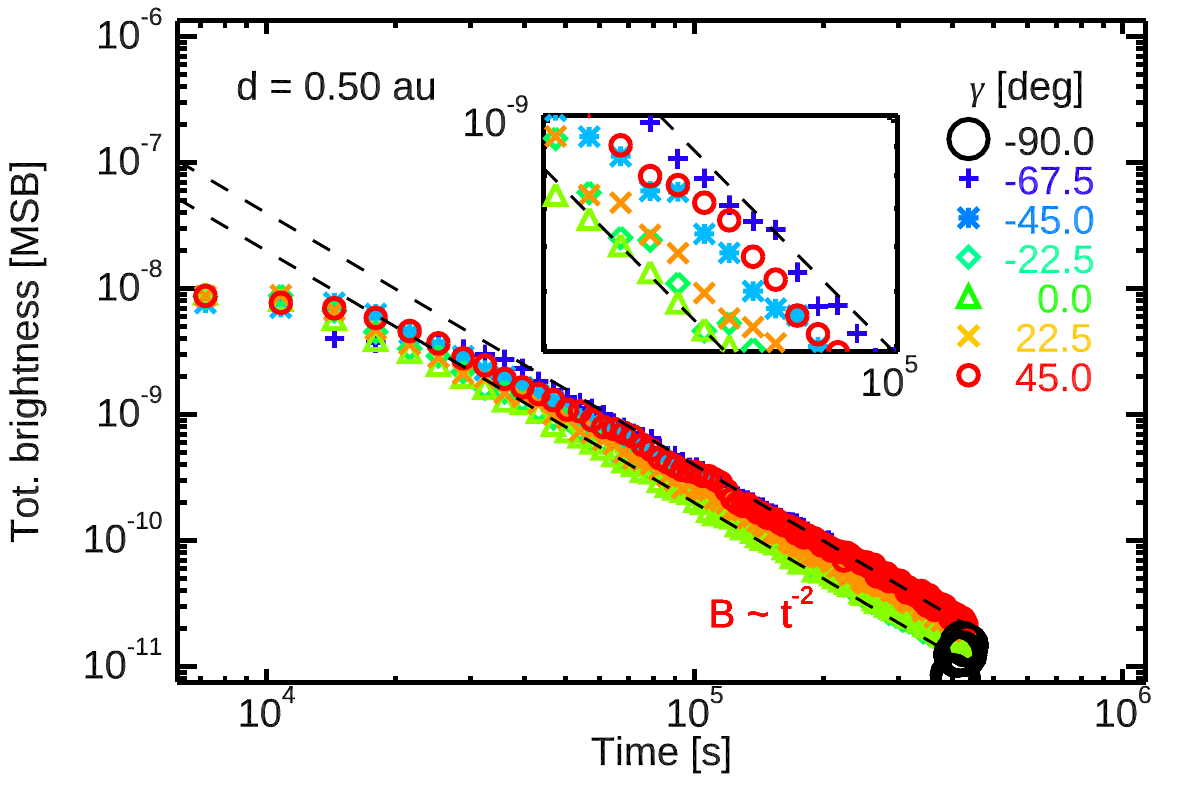} &
			\includegraphics[trim=3 2 1 2,clip,width=0.465\textwidth]{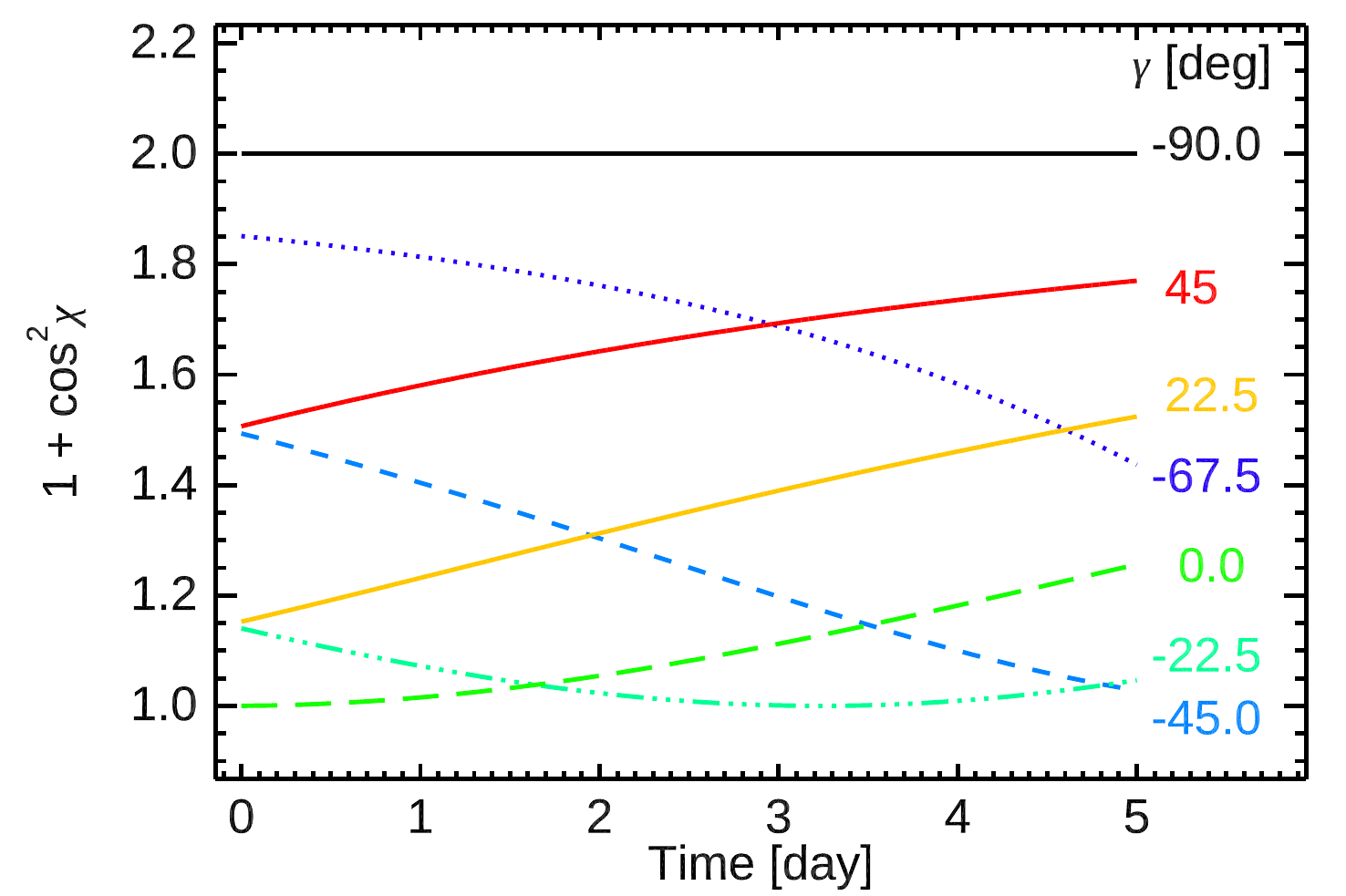} \\	
		\end{tabular}
		\caption{Average total brightness profiles vs. time in log-log scale for a sphere moving along different heliocentric longitudes $\gamma$ and with speed of 100 km s$^{-1}$, observed from a distance of 0.50 au (left). Associated temporal trends of the quantity $1+\cos^2\chi$ for the analysed cases.}
		\label{fig_fo_long}
	\end{figure*}

We also performed some simulations with the density sphere launched at different heliocentric longitudes (or exit angles), spanning from $-90.0 \deg$ (Sun-observer line) to $45.0 \deg$ (behind the plane-of-sky) at steps of $22.5 \deg$. The sphere moves with a constant speed of 100 km s$^{-1}$, and the observer is positioned at 0.5 au from the Sun. The total brightness profiles are shown in the left panel of Fig. \ref{fig_fo_long}. The overall trends follow a power law with index of -2, as expected.

Since the radial distance at any time is the same in all the cases, the magnitude of the total brightness varies because of the scattering angle $\chi$. The data points are closely clustered, but some shift between each series is evident (see the inset plot in the left panel of Fig. \ref{fig_fo_long}): for example, the data points at longitude $\gamma = 0.0 \deg$ (green triangles) lie below the series at $\gamma = $ $22.5 \deg$ (orange crosses) and $45.0 \deg$ (red circles). In the right plot we show the temporal trend for the quantity  $1 + \cos^2\chi$, which determines the difference in the total brightness profiles.  Doing the same comparison, we notice that the curve for $\gamma = 0.0 \deg$ lies below that one at $22.5 \deg$ and $45.0 \deg$. A similar argument holds for any other curve. However, when the blob moves along the Sun-observer line, no total brightness is recorded (black line in the left panel of Fig. \ref{fig_fo_long}), despite the scattering angle $\chi$ being equal to $180.0 \deg$ (hence, the quantity $1 + \cos^2\chi = 2$). This is due to the setup of the raytracing code, which masks the solar disk and does not perform any calculation for LOSs within the solar disk.

\section{Raytracing simulations for PSP/WISPR}
\label{sect4}

Here, we present simulations of density spheres observed by WISPR-I. The simulations are performed for the perihelion transit in 
June 2024, when PSP is expected to reach a minimum heliocentric distance of about $\sim 10~\mathrm{R}_\odot$. The density and size of the sphere are the same as the previously presented cases. We discuss these different scenarios as follows.

\subsection{Observations of a stationary density structure}\label{subs:psp_fb}

\begin{figure*}[htpb]
			\centering
			%\begin{tabular}{c c}
				\includegraphics[trim=3 2 1 2,clip, width=0.6\textwidth]{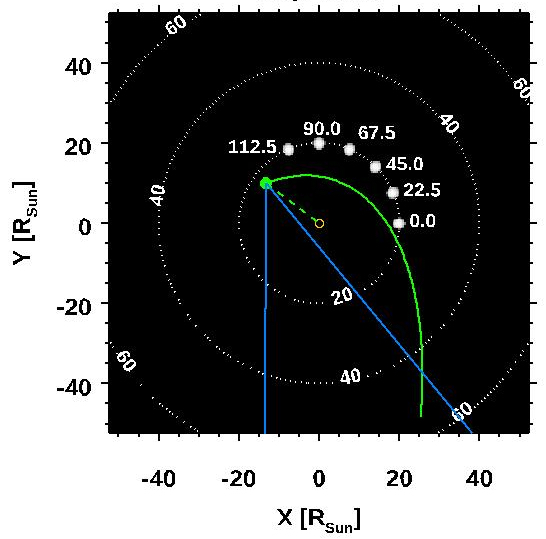}  %&
			%	\includegraphics[trim=3 2 1 2,clip,width=0.53\textwidth]{figure/plot01_050719.eps} \\
			%\end{tabular}
			\caption{Top view of the orbit of PSP and the locations of the density spheres for the studied cases. The PSP trajectory is in green, and the angular WISPR FoV is in blue. The associated animation online is \href{run:movie/movie3.mp4}{movie3.mp4}.}
			\label{fig_psp_fixed}
		\end{figure*}

We simulated a sphere at a constant distance of 20 $\mathrm{R}_\odot$ from the Sun, but positioned at different heliocentric longitudes $\gamma$, between $0.0 \deg$ and $112.5 \deg$, at steps of $22.5 \deg$. A top-view of the orbit and blob locations is given in Fig. \ref{fig_psp_fixed}. The trajectory of PSP is represented by the green curve, with the WISPR angular FoV in blue. The locations of the density structures are represented by the white circles. During the PSP transit, the WISPR FoV intercepts the structures at different times. 

The average total brightness per pixel as a function of time is shown in the right panel of Fig. \ref{fig_psp_fixed} and measured from the starting time $t_0 = $ \lq2024-06-25T00:00:00.000\rq. The total brightness remains almost constant for the structure located at $\gamma=0.0 \deg, 22.5 \deg, \mathrm{and}~45.0 \deg$ (black, pale, and green lines, respectively). The blob-observer distance is continuously changing but this does not affect, as expected, the measured total brightness. However, these profiles do not coincide but are offset between each other, with an increase of the average value with the heliocentric longitude. At longitudes of $67.5 \deg$, $90.0 \deg$ and $112.5 \deg$, the total brightness is measured at later times (the structures enter later in the FoV of the instrument), exhibiting a slow declining trend. The reasons of these offsets and the lack of a constant profile depend on the change in the scattering angle $\chi$. 

\begin{figure*}[htpb]
			\centering
			\begin{tabular}{c c}
				\includegraphics[trim=2 3 3 5,clip,width=0.465\textwidth]{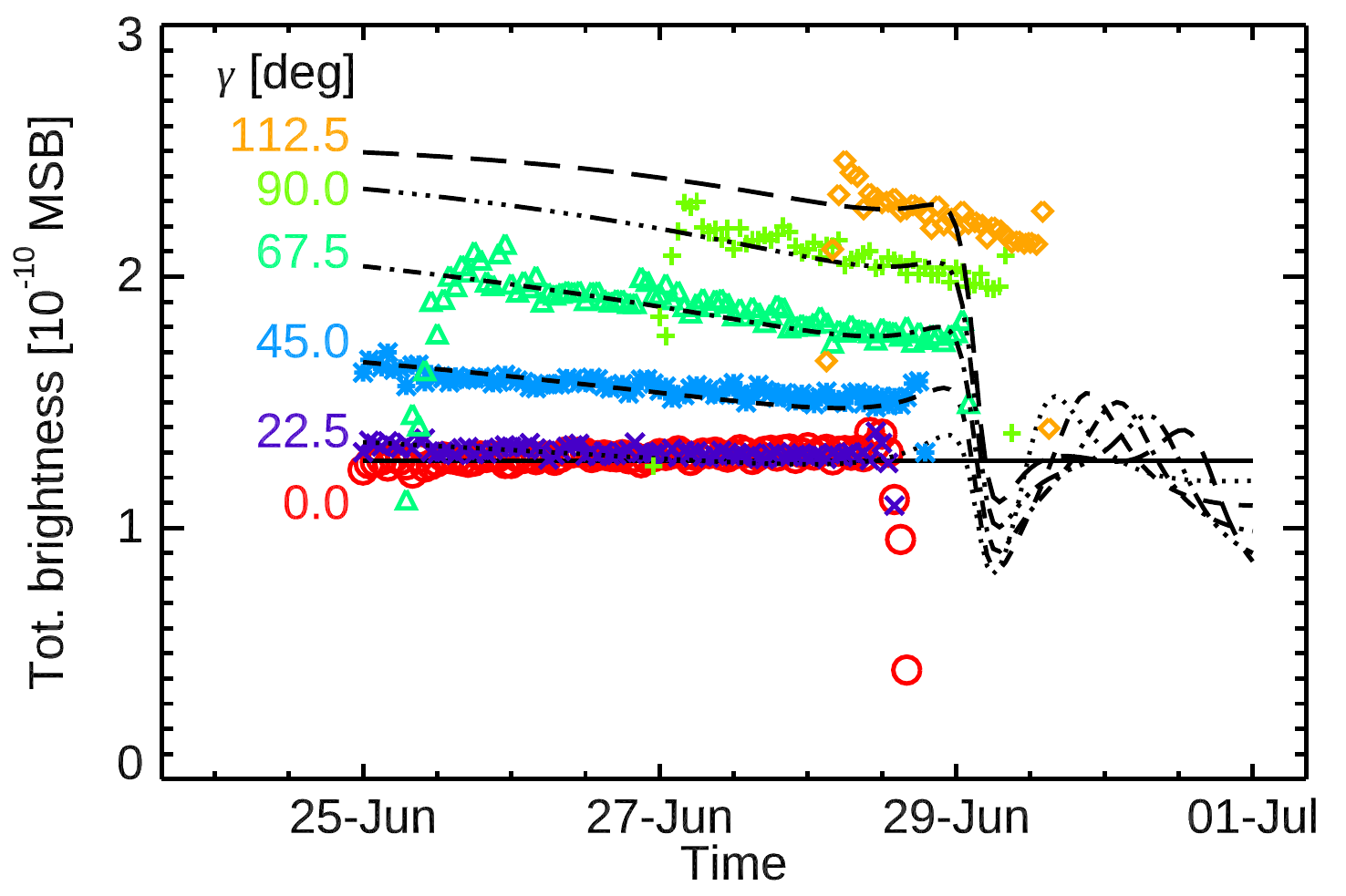} & 
				\includegraphics[trim=4 3 3 5,clip,width=0.465\textwidth]{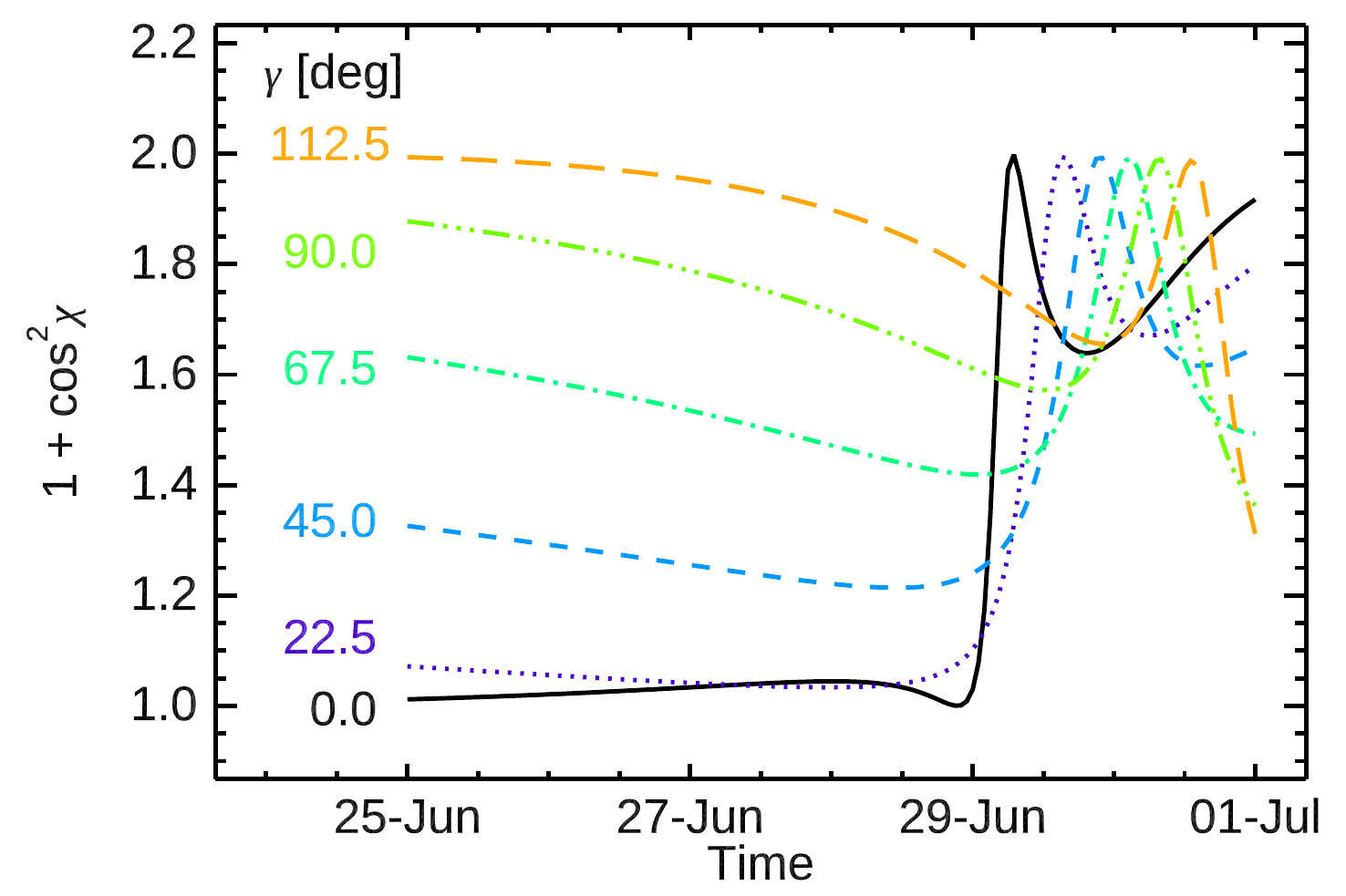} \\
			\end{tabular}
			\caption{Average total brightness as a function of time for a plasma sphere located at a distance of 20 $\mathrm{R}_\odot$ and at different heliocentric longitudes (left panel). The black lines are fits based on Eq. \ref{fit_chi} and the temporal profiles of $1+\cos\chi^2$  are given in the right panel.}
			\label{fig_psp_fixed1}
		\end{figure*}

In the left panel of Fig. \ref{fig_psp_fixed1} we plot the associated quantity $1 + \cos^2 \chi$ as a function of time for our simulations by considering the position of the blob and the S/C. For the density sphere at $\gamma=0.0 \deg$, $1 + \cos^2\chi$ is almost equal to 1, meaning that the $\chi$ angle is close to $90.0 \deg$ (the structure almost lies on the TS). In the other cases, as we have verified, it is $\chi <  90.0 \deg$ , i.e. the structure is out of the TS. In addition, the average total brightness is larger than that for $\gamma=0.0 \deg$. We manually fitted the total brightness profiles by recalling the function in Eq. \ref{eq_tb} and  by considering as a reference the quasi-constant profile at $\gamma=0.0 \deg$. Since the radial distance $r$, the number density $n$, and the column depth $h$ are approximately constant, the only free parameter is the scattering angle $\chi$. The first profile is manually fitted by a constant function $\tilde{B}_{\gamma=0} = 1.3 \times 10^{-10}$ MBS. 
%Given $r_0$, $n_0$ and $h\approx 2 R_\odot$, we can calculate the value of $k$ in Eq. \ref{eq_tb}, which results to be $1.1\times 10^{-14} \mathrm{MSB~m^4}$.
Therefore, the associated scattering angle is assumed to be $\chi_{\gamma=0}=90.0 \deg$. The other profiles are fitted by the function 
	\begin{equation}
			B_{\gamma}(t) = \tilde{B}_{\gamma=0} \left(1+\cos\chi^2_{\gamma}(t)\right) 	
	\label{fit_chi}
	\end{equation} 
	with $1 + \cos\chi^2_{\gamma}(t)$ the profiles in the right panel of Fig. \ref{fig_psp_fixed1} (right panel). In the left panel of Fig. \ref{fig_psp_fixed1}, we note that total brightness profiles are remarkably well-fitted by the black curves obtained from Eq. \ref{fit_chi}.  
%The early fall of the fitting profiles at $\gamma$= 90, 112.5 deg is caused by the finite size of the density sphere ($R = 1 R_\odot$). Indeed, we are dealing with average brightness per pixel, hence reducing the object {\bf as} a point-like source.	      

This outcome confirms the conclusions presented in \citet{Howard2012}, i.e., for features located at the same distance from the Sun but at different heliocentric longitudes,  the resulting scattered emission is minimum for those objects located on the Thomson sphere.
				  
\subsection{Density sphere moving with constant speed}

		\begin{figure}[htpb]
			\centering
			\begin{tabular}{c}
				\includegraphics[width=0.7\textwidth]{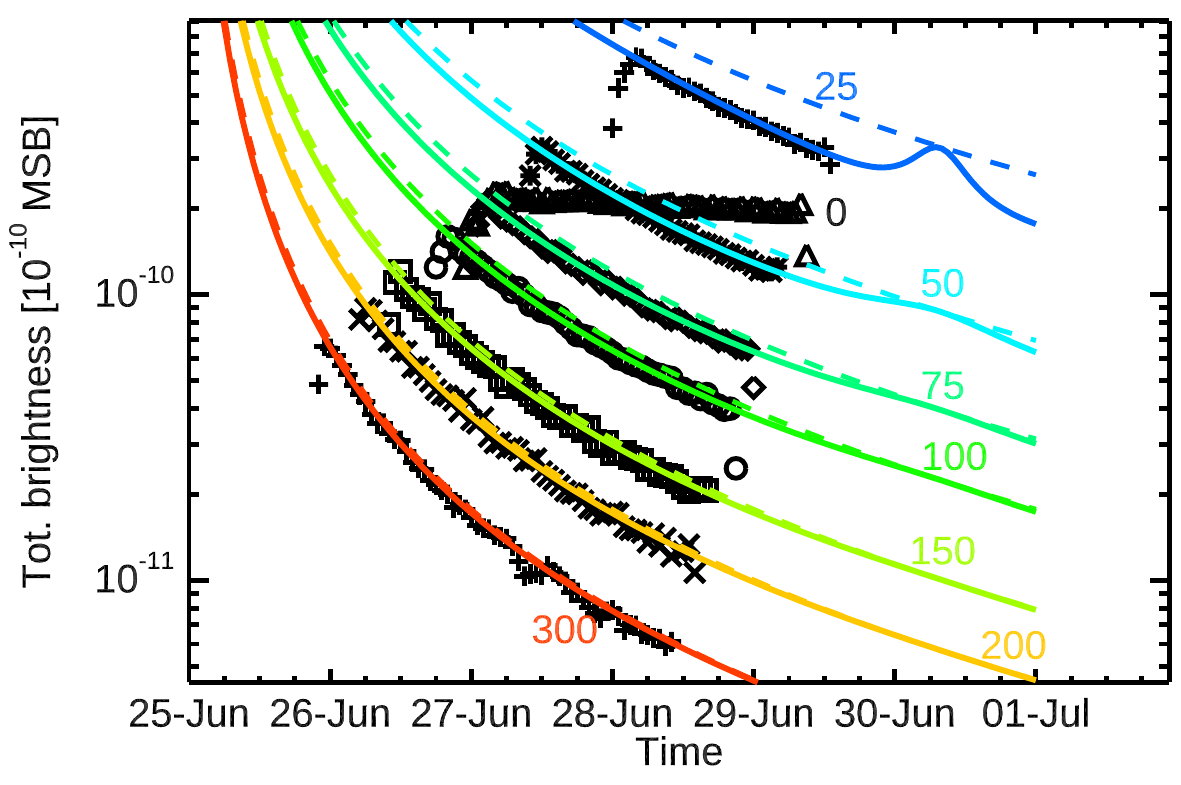} \\ 
				\includegraphics[width=0.7\textwidth]{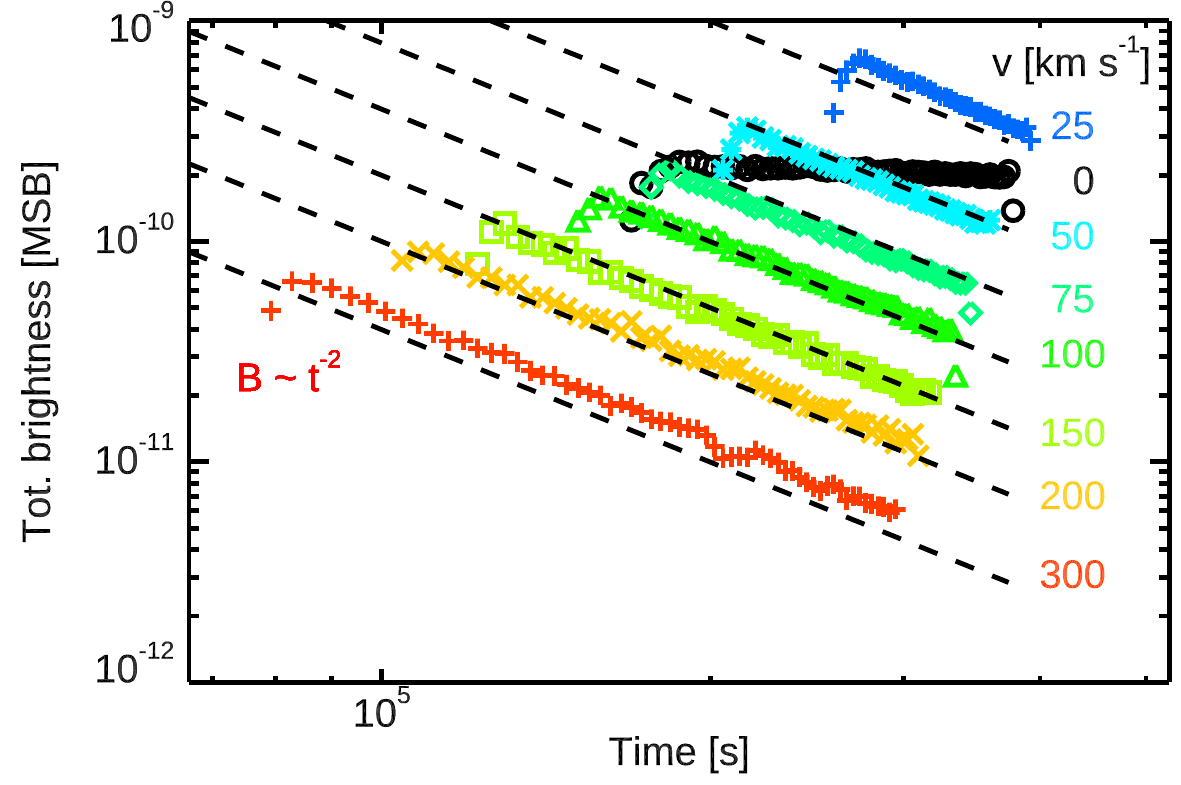} \\
					\includegraphics[width=.7\textwidth]{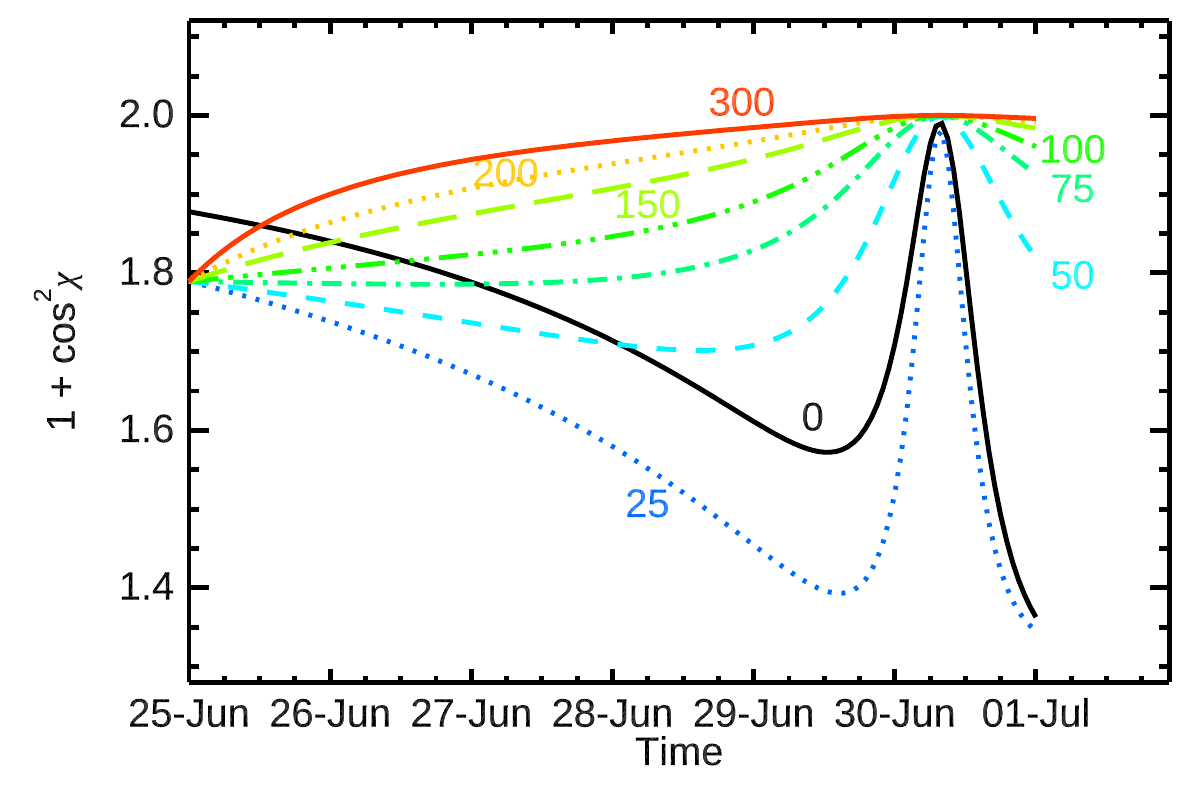} \\
				\end{tabular}
				\caption{Average total brightness vs. time measured by WISPR for a blob moving with different speeds in semi-log (top) and log-log (middle) scales. Variation of the scattering angle function over time (bottom).}
				\label{fig_psp_speed}
		\end{figure}

In addition, we studied the case for a density sphere moving with constant speed along the positive direction of the $y$-axis of the HCI system, i.e., at $\gamma=90.0 \deg$. Here, we used different values of speed while the feature keeps its size and density constant. An example is given in the animation \href{run:movie/movie4.mp4}{movie4.mp4}.
We considered the average total brightness per pixel over time, which is plotted in the top and middle panels of Fig. \ref{fig_psp_speed}. Since the distance of the sphere changes evenly with time, we expect the total brightness to scale as $t^{-2}$, as demonstrated in the log-log scale plot in the middle panel of Fig. \ref{fig_psp_speed}.
Therefore, under the approximation of constant size and density for the structure, such scaling could indicate the propagation of a structure with a constant velocity. Initially, we neglected the factor $1+\cos\chi^2_v(t)$ and used instead the function 
	      \begin{equation}
		      \tilde{B}_{v}(t) = \frac{C} {[r_0 + v (t-t_0)]^2}.
		      \label{eq_approx_b}
	      \end{equation}
		      
Starting from the lowest profile with $v=300$ km s$^{-1}$,  we 
%visually 
fitted the total brightness profiles with the following parameters' values: $C= 4.9\times10^{10}~\mathrm{MSB~m^2 }$, $t_0=$ \lq2024-06-25T00:00:00.000\rq, $r_0 = 1.0~\mathrm{R}_\odot$  and $v$ as the input values in our raytracing simulations. We found that the theoretical curves (coloured dashed lines) and the measured total brightness profiles (black) in the top panel of Fig.~\ref{fig_psp_speed} remarkably coincide for speeds greater than 100 km s$^{-1}$. For speeds below 100~km s$^{-1}$, the structure is imaged by WISPR when PSP is relatively close to it, so there exists a non-negligible contribution from the factor $1+\cos^2\chi_v(t)$. 
		      
To better understand the behaviour of the emission profiles, we considered the scattering angle $\chi_v(t)$ for the different values of speed. As in the previous sections, in the bottom panel of Fig. \ref{fig_psp_speed} we show the scattering angle factor as a function of time. All the scattering angles are less than $90 \deg$, which indicates that the observed structures are out of the TS, and the total brightness decreases with the increase of the speed (hence the distance of the feature from the Sun). 
By considering the brightness profile of the sphere given by Eq. \ref{eq_approx_b} with $v=300$ km s$^{-1}$as a reference, we implicitly assumed $\chi_{v=300} = 0 \deg$ (hence, $1+\cos^2\chi_{v=300}=2$). Therefore, to correct the fit of the profiles we used then the function given by

	      \begin{equation}
		      B_v(t) = \tilde{B}_{v=300}(t)\frac{1+\cos^2\chi_v(t)}{2}
	      \end{equation}
The empirical profiles are now well-fitted by the continuous curves (top panel of Fig. \ref{fig_psp_speed}). This demonstrates that the scattering angle is a crucial parameter to be considered in the analysis of the total brightness of the features recorded by the WISPR telescopes.

\section{Geometric triangulation with WISPR}
\label{sect5}

An important task for WISPR is the determination of the 3D location of the observed structure. Standard techniques for geometric triangulation require at least two observers at different vantage locations observing simultaneously (e.g., from both STEREO S/C or from one STEREO S/C and SoHO). However, a second viewpoint to support WISPR observations might not always be available and accessible by remote imagers during PSP's perihelia. On the other hand, PSP is moving so fast that WISPR could observe the same coronal structures from different perspectives before they change during the time of observation, opening the possibility to rotation tomography \citep{Vasquez2019}.

%\begin{figure*}[htpb]
%	\begin{center}
%	%\begin{tabular}{c c}
%	%	\includegraphics[trim=10 10 40 15,clip,width=.465\textwidth]{../figure/aa/Folie4.png} &
%		\includegraphics[trim=10 10 40 15,clip,width=.65\textwidth]{figure/Folie3.png} \\
%	%\end{tabular}
%	\caption{Geometric triangulation with PSP for a structure P propagating outwards with constant speed.}
%	\label{fig_triang}
%	\end{center}
%\end{figure*}

\begin{figure}[htpb]
\begin{center}
	\begin{tabular}{c}
		\includegraphics[trim=1 2 2 2,clip,width=0.8\textwidth]{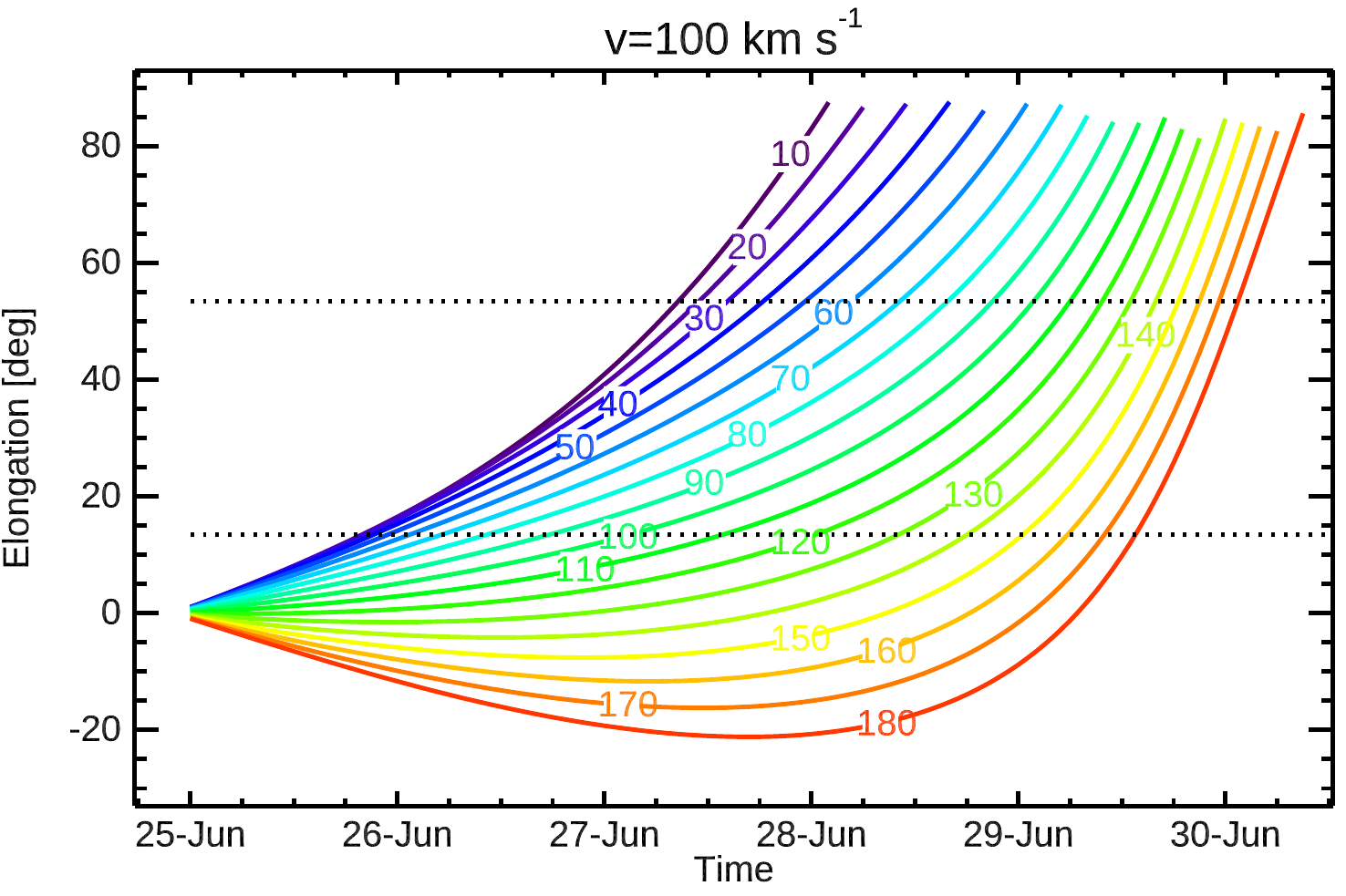} \\ 
		\includegraphics[trim=1 2 2 2,clip,width=0.8\textwidth]{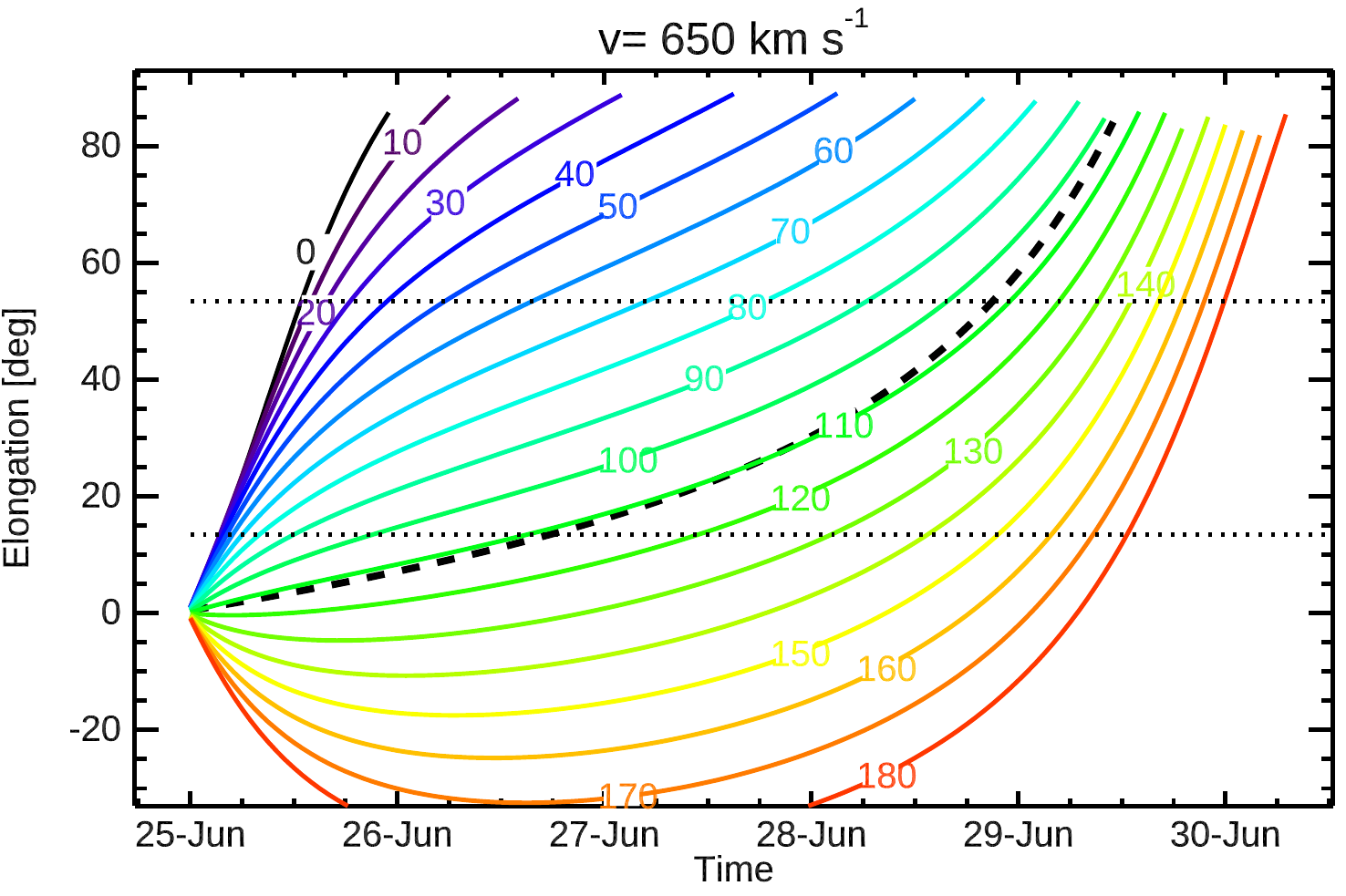} \\
	\end{tabular}
	\caption{Elongation curves for a feature moving at different heliocentric longitudes (coloured labels) and with a speed of 100 km s$^{-1}$ (top) and 650 km s$^{-1}$ (bottom). The black line in the bottom panel represents the curve with $v=100$ km s$^{-1}$ and $\gamma=90 \deg$, which is quasi-identical to that with $v=650~\mathrm{km~s}^{-1}$ and $\gamma=110 \deg$.}
	\label{fig_elong}
	\end{center}
\end{figure}

The basics for geometric triangulation with WISPR are presented and discussed in \citet{Liewer2019} for the case of a structure located in and out of PSP's orbital plane. Here, we recall the former case.
In Figure \ref{fig_sketch} we showed a scheme of geometric triangulation for a feature P lying in the orbital plane of PSP and moving away with a constant speed. The position of P is determined in polar coordinates by the pair $(r,\gamma)$. The elongation $\xi$ of P, which is assumed moving with  constant speed $v$, measured with WISPR is

\begin{equation}
	\xi(t) = \tan^{-1}\left\{\frac{[r_0 + v (t-t_0)]\sin(\gamma - \alpha(t))}{d(t) - [r_0 + v (t-t_0)] \cos(\gamma - \alpha(t))}\right\},
	\label{eq_elong}
\end{equation}
with $d(t)$ the distance of PSP from the Sun, $\alpha(t)$ the azimuthal angle of PSP in a given reference frame (it could simply correspond to the heliocentric longitude angle for PSP in the HCI coordinate system if its orbital plane were the solar equatorial plane).
The time $t$ is measured from a starting time $t_0$. The only free parameters are  $v$ and $\gamma$. In fact, the distance $r_0$ at time $t_0$ can be expressed in terms of $\gamma$ as $r_0 = d(t_0) \sin\xi(t_0)/\sin(\gamma+\alpha(t_0)+\xi(t_0))$.  Fitting analysis of elongation measurements vs. time (e.g., via standard routines like MPFIT, which is available in IDL/SSW) can be used to estimate the values of the parameters. However, triangulation of moving features in real WISPR data is not an easy task, since the convergence to a solution in the fitting algorithm is highly dependant on the initial guessed values of the parameters. 
For example, on the top panel of Fig. \ref{fig_elong} we show a set of theoretical elongation curves as a function of time for a feature moving along different heliocentric longitudes with the same speed of 100~km~s$^{-1}$. The starting time $t_0$ is \lq2024-06-25T00:00:00.000\rq and the starting height, $r_0 = 1~\mathrm{R}_\odot$. If we consider the curve corresponding to a longitude of $90 \deg$, we notice that it almost coincides with that having a speed of 650~ km~s$^{-1}$ and longitude of $110 \deg$ (black line in Fig. \ref{fig_elong}-bottom). Since there is a small separation between the two curves, the degeneracy might disappear at higher values of elongations angles (e.g., in the FoV of WISPR-O). However, in a height-time map (hereafter J-map)
%if one studies 
the corresponding tracks would be practically identical because of their thickness and errors associated with the sampling.

The analysis of the total brightness can improve the estimates of the unknown parameters. Indeed, if we recall the definition in Eq. \ref{eq_tb}, we notice that the total brightness contains information about the spatial location of a feature via the distance $r$ and the scattering angle $\chi$, which implicitly depend on $\xi$ and $\gamma$. 

\subsection{Case Studies}

\begin{figure*}[htpb]
	\begin{tabular}{c}
	\includegraphics[trim=2 0 8 6,clip, width=0.9\textwidth]{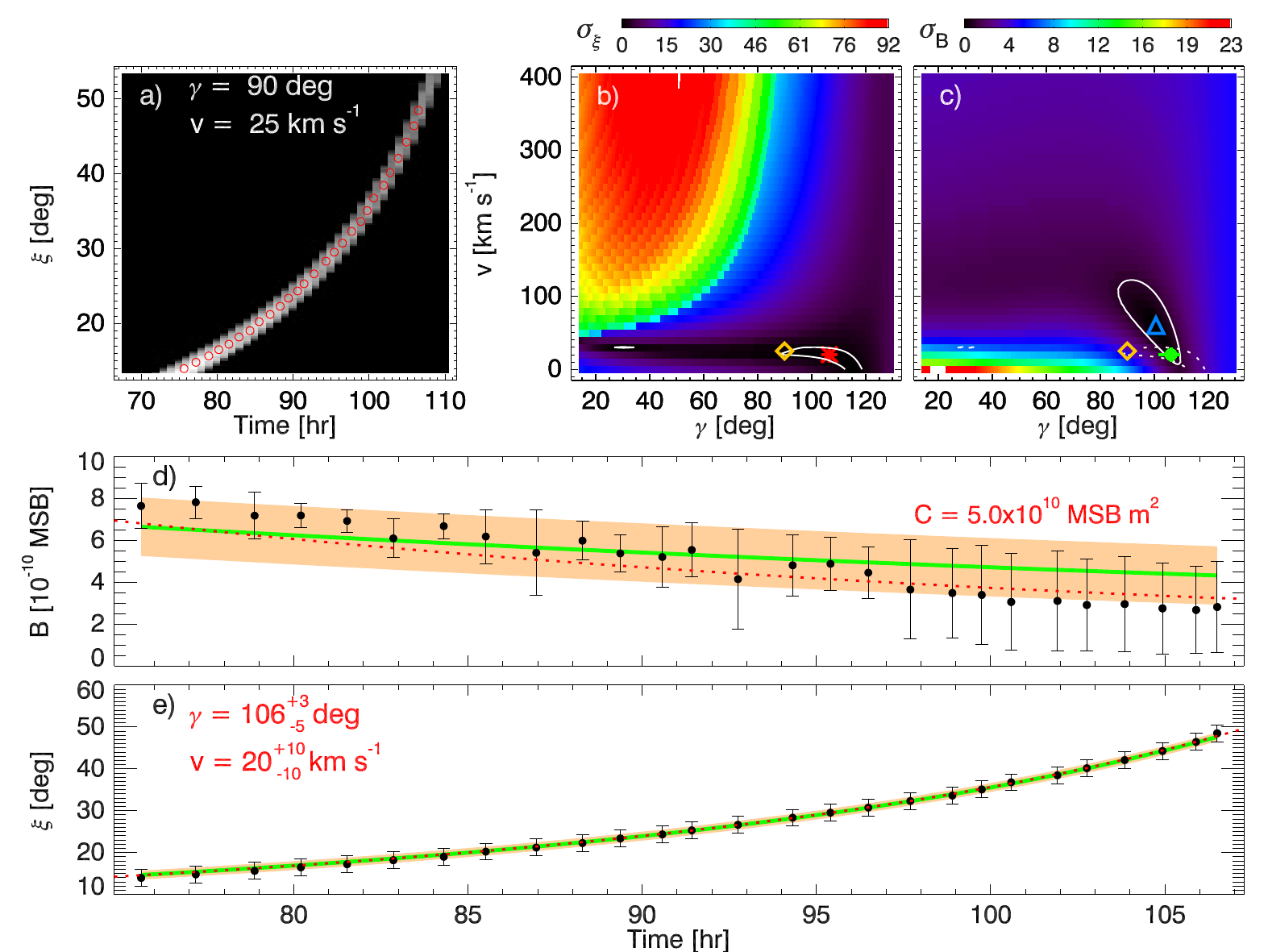} \\
	\includegraphics[trim=2 0 5 2,clip,width=0.9\textwidth]{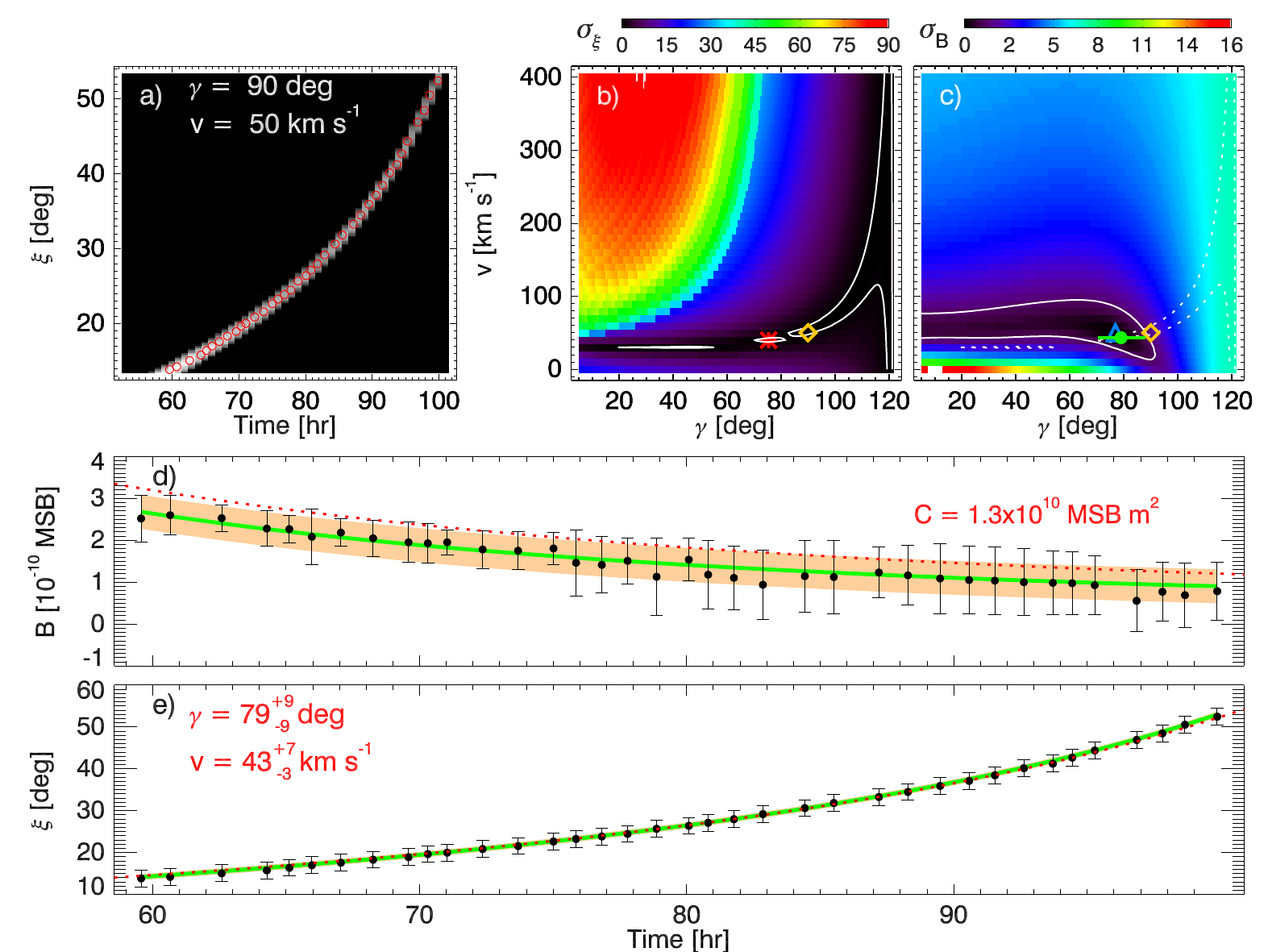} \\
	\end{tabular}
	\caption{\small{Fitting analysis for two cases of a density sphere moving with $\gamma = 90 \deg$ and $v = 25~\mathrm{and}~50~\mathrm{km~s}^{-1}$, respectively . Panel {\it a)}: a J-map from the synthetic images showing the signature of the propagating plasma sphere (the values of the simulations parameters are reported at the top of the map). The red circles mark the data points manually determined and used to extract samples of the total brightness vs. time and elongation vs. time. Panels {\it b)} and {\it c)}: $\sigma_\xi$ and $\sigma_B$ maps. The white contour outlines the 3$\sigma$ region. The red asterisk and the blue triangle mark the global minimum in each map, respectively. The true values of the parameters is marked by the yellow diamond, while the best fit model is returned by the green dot and overplotted in the $\sigma_B$ map. Panels {\it d)} and {\it e)}: fits of the total brightness and elongation measurements. The data points with associated error bars are in black, the fitting lines are given in green. The dashed red lines are the exact solutions for the total brightness and the elongation (obtained with Eq. \ref{eq_tb} and \ref{eq_elong}). The shaded areas represent the 3$\sigma$ confidence interval. For the elongation measurements, the confidence interval is very narrow and almost hidden by the fitting line. The estimates of the parameters are reported in red.}}
	\label{example}
\end{figure*}

\begin{figure*}[htpb]
	\begin{tabular}{c}
	\includegraphics[trim=2 0 8 6,clip, width=0.9\textwidth]{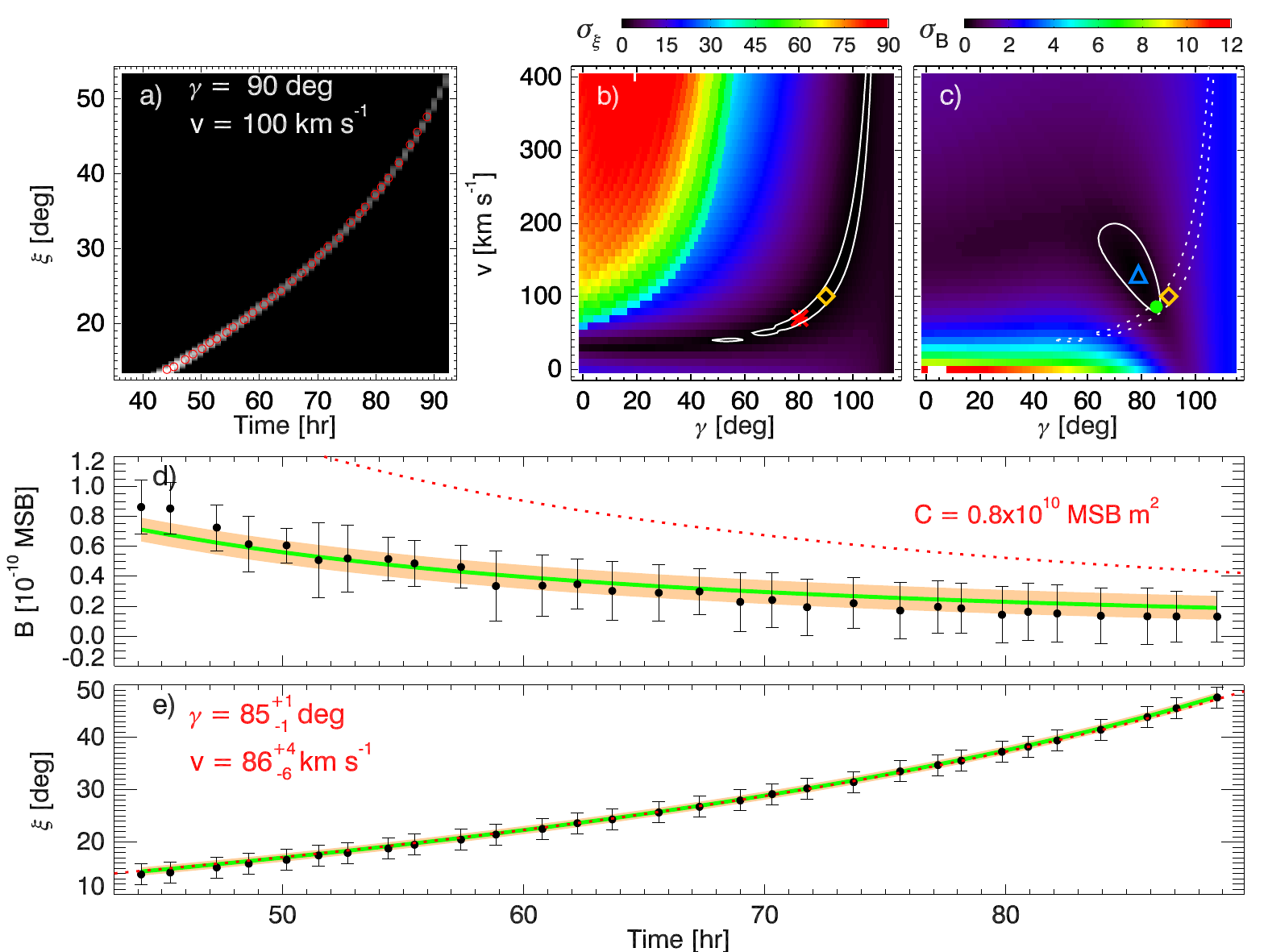} \\
	\includegraphics[trim=2 0 5 2,clip,width=0.9\textwidth]{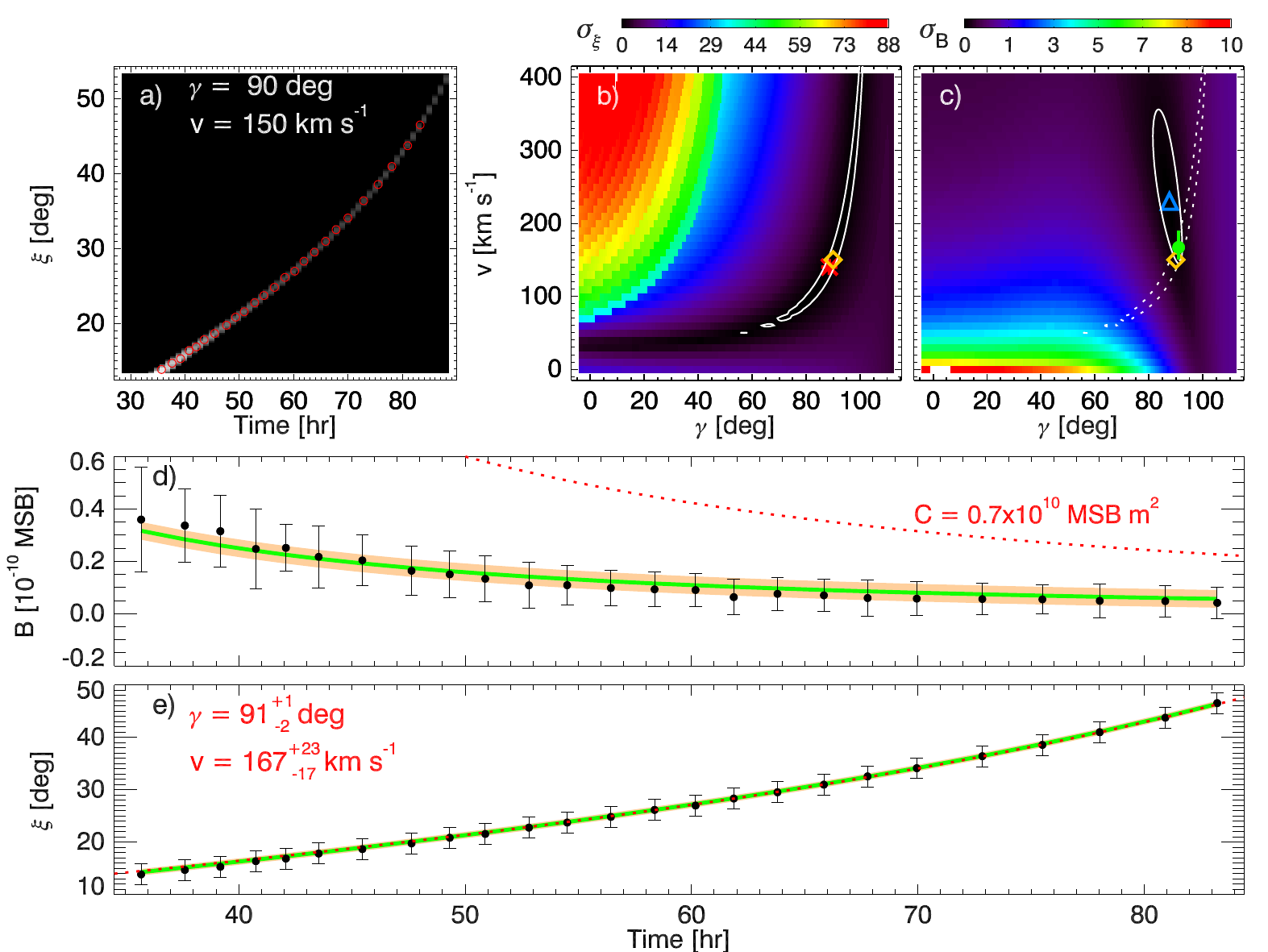} \\
	\end{tabular}
	\caption{The same as in Fig. \ref{example} for a density sphere with $\gamma=90$ deg and $v=100,150$ km s$^{-1}$.}
	\label{example1}
\end{figure*}

We studied the cases for a spherical plasma structure moving with speed of 25, 50, 100, and 150 km s$^{-1}$ at a longitude of $\gamma=90 \deg$. To this aim, we created the J-maps displayed in the panels $\it a)$ of Figures \ref{example} and \ref{example1} from synthetic images by considering a horizontal slit 11 pixel wide, passing through the centre of the images. Briefly, we first averaged the brightness over the slit width, and stacked all the resulting columns in a map where the horizontal axis represents the time and the vertical one the elongation distance. Then, we sampled the tracks manually to get measurements of the elongations $\xi_i$ {\it vs.} time (red data points in the J-maps of Fig.~\ref{example}). We assumed an error of about $2 \deg$ in elongation for every measurement. This choice was based on studies of error analysis in data from the STEREO/SECCHI heliospheric imagers (SECCHI/HI) \citep{Williams2009, Barnard2015} and in consideration of the low spatial resolution of our simulations (ten times smaller than the resolution of real WISPR data). Then, for any data point of the J-map we extracted the average total brightness $b_i$ by taking a 2-D boxcar with a size of three pixels. The associated uncertainty for the total brightness is the standard deviation calculated over the boxcar.

To fit the profiles, we adopted the fitting method described in \citet{Rouillard2010}, which was applied to elongation measurements alone of features observed by SECCHI/HI. The idea is to determine the best fit by varying systematically the parameter values and to create different realisation of the model. The comparison between the model and the data measurements is achieved by calculating the standard deviations ($\sigma$), which can be organised in a 2D map ($\sigma$-map) with the axes representing the varying parameters. In \citet{Rouillard2010} the global minimum in the $\sigma$-map defines the best fit model, hence the best parameter values.

In order to consider both the elongations and the total brightness measurements, we had to follow a slightly different approach. We varied the parameters on a large interval of values at steps of $0.5 \deg$ in longitude $\gamma$ and 10 km s$^{-1}$ in speed $v$. The coefficient $C$ for the total brightness ranges between $[0.1 - 5.0]\times10^{10}$ MSB m$^2$ in steps of 0.1$\times10^{10}$ MSB m$^2$. The panels {\it b)} and {\it c)} on Fig. \ref{example} and \ref{example1} show the maps for the elongation $\sigma_\xi$ (middle) and the total brightness $\sigma_B$ (right) as a function of $\gamma$ on the horizontal axis and $v$ on the vertical one. We considered a region of 3$\sigma_\xi$ around the global minimum (marked by a red asterisk). The true value of $\gamma$ and $v$ is marked with a yellow diamond. It can be noticed that these two points are close to each other, but in general they do not coincide. The 3$\sigma_\xi$ region extends for large values of longitude and speed. It is therefore necessary to narrow this area around the best value of the parameters. Hence, since we have three free parameters ($C, \gamma, v$), we can build a datacube of the total brightness standard deviation. Then, for any value of $C$, i.e., for any cut of the datacube, we looked for the global minimum of each resulting 2D map (blue triangle) and the associated 3$\sigma_B$ region.

The right panel on Fig. 10 shows a cut of the datacube for the best $C$. The best value of $C$ is chosen when the global minimum in the $\sigma_B$-map is the closest one to that of the $\sigma_\xi$-map and the overlap between the two regions is not null. The best longitude and speed are found as the centroid of the intersecting area with errors obtained from the extension of this area (green dot with some vertical and horizontal bar, sometimes hidden by the size of the dot). Because of the irregular shape of the area, the errors around the centroid can be asymmetric. We did not explicit error bars for $C$, but this can be assumed equal to the grid step (i.e., 0.1$\times10^{10}$ MSB m$^2$).

The bottom panels {\it d)} and {\it e)} show the data point measurements of the total brightness and elongation vs. time from the J-map fitted with a curve with the parameter estimates indicated in red. Besides the fits (green lines), we plot the exact solutions for the brightness and elongations (red dashed line), and the $3\sigma$ confidence interval (shaded area), which gives a measure of the uncertainty of the fitting model. The value of $C$ is different in the different analysed cases, despite the simulations are indeed performed for a sphere with the same density and size.

As already mentioned, $C$ is a measure of the density along the column depth. Since the total brightness measurements are taken directly from the J-map, which was built from a slit with a finite width, and hence does not cover the entire blob, they may be smaller than the \lq\lq true\rq\rq brightness shown in Fig. 8 (which was calculated by averaging over the apparent area of the feature). Moreover, the values of $C$ are not random but decrease as the speed of the blob increases (an effect related to the relative distance between the blob and the PSP S/C).

For better estimates and a proper assessment of the density and mass of the structure, it would be necessary to calculate the total brightness over the total apparent area of the structure as seen in the WISPR FoV. Here, we simply demonstrate that the analysis of the elongation and brightness profiles from a single J-map can provide reliable estimates, sufficiently close to the ``true'' values, of the speed and position for a propagating structure.

To minimise Eqs. \ref{eq_tb} and \ref{eq_elong} simultaneously, further fitting approaches could be implemented, e.g., Montecarlo, Bayesian techniques \citep[][]{Pascoe2017, Sharma2017} but that is beyond the scope of the present paper.

%%%%%%%%%%%%%%%%%%%%%%%%%%%%%%%%%%
\section{Conclusion} 
\label{sect6} 

Compared to past coronagraphic observations (e.g., from SoHo and STEREO), image data from the WISPR instruments will require novel techniques of analysis and particular care in the interpretation of the data because of the continuous changes of the plane-of-sky and distance of PSP from the Sun. In this paper, we presented a preliminary study on the total brightness variation of a simulated coronal structure on the solar equatorial plane as it would be observed by the WISPR inner telescope during a future PSP perihelion transit in June 2024. 

For simplicity, we relied on the approximation that the orbital plane of PSP coincides with the solar equatorial plane, therefore reducing our analysis to a simple 2D geometry. To create synthetic images of WISPR, i.e., images of the total brightness emission due to Thomson scattering, we used the raytracing software \citep{Thernisien2006}, adapted to the orbit of PSP and to the FoV of WISPR. We considered a very simple structure, namely a density sphere with a radius of 1~$\mathrm{R}_\odot$, which is intended to mimic the appearance of either a simple CME or blob in the WISPR FoV. 

To better understand the effects of the effects of the varying distance between the observer and the coronal feature, we have taken into account two cases: a stationary sphere in a fixed position and a sphere propagating outwards from the Sun with different constant speeds. The space-time evolution of the structure is determined by the radial speed $v$ and the heliocentric longitude $\gamma$. We determined the mathematical expressions for the total brightness and the elongation measured by WISPR (Eq. \ref{eq_tb} and \ref{eq_elong}) as a function of time with $C, \gamma,~\mathrm{and}~v$ as free parameters.

As expected,  we found that the time profiles of the total brightness depend on both the radial distance of the structure from the Sun and the scattering angle. In the case of the moving plasma spheres, we notice that the dependence on the scattering angles tends to attenuate when the structure is located at large distances from the Sun. This implies that the structure is imaged by WISPR when the distance between the structure and PSP is large or, in other words, the structure is located far from the Thomson sphere. In turn, this can occur when the structure has a speed typically larger than $100~\mathrm{km~s}^{-1}$, so that it can move far in a shorter time. However, we have not properly studied the dependence on the heliocentric longitude for the orbit of PSP for this case. 

These findings suggest that the total brightness evolution could be exploited to obtain a more precise triangulation of the observed features and to remove the degeneracy that affects the elongation measurements in J-maps. Indeed, we have demonstrated that distinct set of values for the free parameters $(\gamma, v)$ may be associated with quasi-identical time patterns of the elongation. Therefore, our analysis suggests that a fitting procedure involving both total brightness and elongation measurements can improve the estimates of the parameters. For this task, we have used here a minimisation technique adapted from \citet{Rouillard2010}. We found that the estimates of the longitude $\gamma$ and the propagation speed $v$ are good, with deviation from the ``true'' values used in the simulations between 1 and 20 $\%$. 

Notice that the outcomes of this work are also relevant for the analysis of other coronal features like propagating blobs \citep{Viall2015}, equatorial coronal hole jets \citep{Nistico2010, Roberts2018} or wave disturbances in coronal funnels \citep{Nistico2014, Goddard2016}. The results also apply to SoHO and STEREO white-light observations as originally suggested by \citet{Vourlidas2006}. Here, we have not studied, however, the effects of the overlapping of different coronal structures, features extended along the LoS, or other expansion effects that may affect the brightness profiles.  

      %%%%%%%%%%%%%%%%%%%%%%%%%%%%%%%%%%%%%%%%%%%%%%%%%%%%%%%%%%%%%%%%%%%%%%%%%%%
\section*{Acknowledgment}
 The authors acknowledge the WISPR team throughout this work. GN and VB acknowledge the support of the Coronagraphic German and US SolarProbePlus Survey (CGAUSS) project for WISPR by the German Aerospace Centre (DLR) under grant 50OL1901 as a national contribution to the Parker Solar Probe mission. AV is supported by WISPR task at the Applied Physics Laboratory of the Johns Hopkins University. The work of PCL was conducted at the Jet Propulsion Laboratory, California Institute of Technology under a contract from NASA. AT, GS, RH acknowledge the support of the NASA Parker Solar Probe Program.

%\acknowledgment US spelling: \verb+\acknowledgment+
%\acknowledgement British  spelling: \verb+\acknowledgement+

     % format of references provided by the journal (.bst)
\bibliographystyle{plainnat}
     % name your Bibtex file containing your references (.bib)
\bibliography{references}  
 
\end{document}